\documentclass[a4paper,11pt]{article}

\usepackage{latexsym}
\usepackage{amsmath}
\usepackage{a4wide}
\usepackage{graphicx, subfigure}
\usepackage{cite}
\usepackage{tabularx}
\usepackage{array}
\usepackage{graphics}
\usepackage{graphicx}
\usepackage{epsfig}                 
\usepackage{amsmath}
\usepackage{amssymb}
\usepackage{psfrag}
\usepackage{longtable}
\usepackage{verbatim}
\usepackage{cite}
\usepackage{color}

\input{definitions.tex}

\begin{document}

\hfill {\tt CERN-PH-TH/2012-174,MZ-TH/12-22}
\begin{center}

\vspace{3cm}

{\huge\bf   The Minimal Flavour Violation benchmark}

\vspace{0.5cm}

{\huge\bf  in view of the latest LHCb data}
\vspace{2cm}

{\Large\bf  Tobias Hurth\footnote{Email: tobias.hurth@cern.ch}
}

\vspace{0.1cm}
{\Large\it Institute for Physics, Johannes Gutenberg University,\\D-55099 Mainz, Germany}\\

 \vspace{1cm}

{\Large\bf   Farvah Mahmoudi\footnote{Email: mahmoudi@in2p3.fr}                            
}

\vspace{0.3cm}
{\Large\it CERN Theory Division, Physics Department,\\ CH-1211 Geneva 23, Switzerland}\\
{\Large\it Clermont Universit{\'e}, Universit\'e Blaise Pascal, CNRS/IN2P3,\\
LPC, BP 10448, 63000 Clermont-Ferrand, France}\\

\vspace{1cm}

 \end{center}
\vspace{2cm}

\thispagestyle{empty}
{\bf Abstract:} 
We derive the consequences of the MFV hypothesis for $\Delta  F =1$ flavour observables based on the latest LHCb data. 
Any future measurements beyond the MFV  bounds and relations  unambiguously indicate the existence of new flavour structures 
next to the Yukawa couplings of the Standard Model. 
\newpage

\section{Introduction}
The last decade of quark flavour experiments has shown an impressive success  of the simple CKM mechanism 
for flavour mixing and CP violation:   All measurements of rare decays ($\Delta F =1$), of mixing phenomena ($\Delta F=2$),
and  of all CP violating observables at tree and loop level have been consistent  with the Cabibbo-Kobayashi-Maskawa (CKM)
theory of the Standard Model (SM); 
in other words none of the former and present  flavour experiments  including  the first generation of the $B$ factories 
at KEK (Belle experiment at the KEKB $e^+e^-$ collider)~\cite{Belle} and at SLAC (BaBar
experiment at the PEP-II $e^+e^-$ collider)~\cite{Babar},
and the Tevatron $B$ physics programs (CDF~\cite{TevatronB1} and
D0~\cite{TevatronB2} experiments) has found an  unambiguous sign of New Physics (NP). 
Moreover,  all first results based on the high statistics of the LHCb experiment~\cite{LHCb} are again very well in agreement with the 
 CKM theory of the SM.  
Of course  there have been and there are still so-called tensions, anomalies, or puzzles in the quark flavour data at 1-,2-, or 3-$\sigma$ level,
however, until now they all have disappeared after some time when more statistics  had been collected. 

This means that   flavour-violating processes between quarks are governed by a $3 \times 3$ unitarity matrix
referred to as the CKM
matrix~\cite{Kobayashi:1973fv,Cabibbo:1963yz}. 
In particular, the one phase among the four real independent parameters of the CKM matrix 
represents the dominating  source of CP
violation and it  allows for a unified description of all the CP violating
phenomena in the SM.  This success of the  CKM theory of CP violation 
was honored with  the Nobel Prize in Physics in 2008.

This feature is somehow unexpected because in principle 
(loop-induced) flavour changing neutral current (FCNC) processes  like  $\bar B \to X_s  \gamma$
offer high sensitivity to NP;  additional contributions to the
decay rate, in which SM particles are replaced by new particles such as
the supersymmetric charginos or gluinos, are not suppressed by the
factor $\alpha/4\pi$ relative to the SM contribution. Thus,  FCNC
decays provide information about the SM and its extensions via virtual
effects to scales presently not accessible (for reviews see Refs.~\cite{Hurth:2010tk,Hurth:2003vb}).
This is complementary to the
direct production of new particles at collider experiments~\cite{Mahmoudi:2012uk, Hurth:2011zy}

Within this indirect search for  NP  there is a an ambiguity of the NP   scale. In the model-independent approach using 
the effective electroweak  hamiltonian, the contribution to one specific operator ${\cal O}_i$ can be parametrized via  
$( C^i_{\rm SM}\, / \,  M_W  +  C^i_{\rm NP}\, / \, \Lambda_{\rm NP} ) \times {\cal O}_i$
where the first term represents the SM contribution at the electroweak scale $M_W$ and the second one the NP 
contribution with an unknown coupling 
$C^i_{\rm SM}$ and an unknown  NP scale $\Lambda_{\rm NP}$.  
The non-existence of large NP effects in  FCNC  observables in general
 implies the
infamous flavour problem, namely why FCNC are suppressed:
Either the mass scale of the new degrees of freedom $\Lambda_{\rm NP}$  is very high or the 
new flavour-violating couplings $C^i_{\rm NP}$  are small for (symmetry?) reasons that remain to be found. 
For example,  assuming  {\it generic\/}  new flavour-violating couplings of $O(1)$,  
the present data on  $K$-$\bar K$ mixing implies  a very high NP scale of order $10^3$--$10^4$ TeV 
depending on whether the new  contributions enter at loop- or at tree-level.   
 In contrast, theoretical  considerations on the Higgs sector, which is responsible for the mass generation 
of the fundamental particles in the SM, call for NP at
order $1$ TeV.    As a consequence,
any NP below the  $1$ TeV scale must have a non-generic flavour structure.

The hypothesis of minimal flavour 
violation (MFV)~\cite{Chivukula:1987py,Hall:1990ac,D'Ambrosio:2002ex}, 
is a formal model-independent  solution to the 
NP flavour problem. It assumes that the flavour and the CP 
symmetry are  broken as in the SM.  Thus, it 
requires that all flavour- and CP-violating interactions be 
linked to the known structure of Yukawa couplings.     
A renormalization-group  invariant definition of MFV 
based on a symmetry principle 
is given in Ref.~\cite{D'Ambrosio:2002ex}; 
this  is mandatory for  a consistent  
effective field theoretical analysis
of NP  effects.

The MFV hypothesis is far from being verified. There is still room for sizable new effects, 
and  new flavour structures beyond the Yukawa couplings are still compatible 
with the present data    because the flavour sector has been tested only 
at the $10\%$  level  especially in the $b\to s$  transitions.
However, the MFV hypothesis represents an  important benchmark in the sense that any measurement  which 
is inconsistent with the general constraints and relations induced by the MFV 
hypothesis  unambiguously indicates the existence of new flavour structures. 

This implies the main purpose of the present paper, namely to derive the  consequences 
of the MFV hypothesis based on the latest LHCb data.  This was done some time ago 
in Ref.~\cite{Hurth:2008jc}  for $\Delta F =1$ observables. In particular the impressive start of the LHCb experiment
suggests to update this analysis. 

Besides the new data from the $B$ factories and from the LHCb experiment we implement  some additional 
theoretical improvements compared to the previous analysis of $\Delta F=1$ processes: The exclusive $B \to K^* \ell^+\ell^-$
decay was  analyzed within a simple form factor analysis. Here we use the up-to-date theoretical tools of QCD-improved 
 factorization for  the low-$q^2$ region~\cite{Beneke:2001at,Beneke:2004dp} and the recently proposed OPE methods for  the 
 high-$q^2$ region~\cite{Grinstein:2004vb, Beylich:2011aq}. Secondly, we skip the approximation that the NP contributions 
 to the electromagnetic and chromomagnetic operators appear in a fixed linear combination which was necessary in the previous analysis
 due to the limited number of independent experimental measurements.

 The MFV analysis of $\Delta F =2$ mixing phenomena decouples from the $\Delta F=1$  analysis within the standard MFV framework.
 The $\Delta F=2$  analysis in Ref.~\cite{Bona:2005eu}  was recently updated in Refs.~\cite{Lenz:2010gu,Lenz:2012az}.

The paper is organized as follows: In Section 2 we recall the definition of the RG invariant definition of the MFV hypothesis and in particular
the effective hamiltonian within this framework. In Section 3 we work out the dependence on the non-standard Wilson coefficients 
of the MFV effective theory 
for all the 
flavour observables used in our analysis. We also discuss the various sources of the  uncertainties in the theoretical
predictions.  In Section 4 we give some numerical details and in Section 5 we discuss  our results.

\section{Effective hamiltonian with MFV}

\subsection{MFV hypothesis}
The SM gauge interactions are universal in quark flavour  space, this means the gauge sector of the SM is invariant 
under the flavour group $G_{\rm flavour}$  which can be decomposed as 

\begin{equation}
G_{\rm flavour} =  U(3)_{Q_L} \times U(3)_{U_R} \times U(3)_{D_R}\,.
\end{equation}

In the SM this symmetry is only broken by the Yukawa couplings. Any new physics model in which all flavour-
and CP-violating  interactions can be linked to the known Yukawa couplings is {\it minimal flavour violating}.   In order to implement 
this principle in a renormalization group invariant way~\cite{D'Ambrosio:2002ex}, one promotes $G_{\rm flavour}$  to a symmetry of the theory by introducing 
auxiliary fields $Y_U$ and $Y_D$ transforming under $SU(3)^3_q$ as

\begin{equation}
Y_U  ~  (3, \bar{3}, 1)\,\,\,  {\rm and}  \,\,\,  Y_D ~ (3, 1, \bar{3})\, .  
\end{equation}

The Yukawa couplings are then introduced as background fields of these so-called spurions transforming under the flavour group. 
An effective theory satisfies the criterion of MFV if all higher-dimensional, constructed from SM and $Y$ fields, are invariant under
CP and under the flavour group $G_{\rm flavour}$~\cite{D'Ambrosio:2002ex}.

In the construction of the effective field theory,  operators with arbitrary 
powers of the dimensionless $Y_{U/D}$ have to be considered in principle. 
However, the specific  structure of the SM,  with  its hierarchy 
of CKM matrix  elements and quark masses,  drastically reduces 
the number  of numerically  relevant operators. For example, it can be shown  
that in MFV models with one Higgs doublet, all FCNC  processes with  external 
$d$-type quarks are governed by 
the following combination of spurions due to the dominance of the top 
Yukawa coupling $y_t$:

\begin{equation}
(Y_U Y_U^\dagger)_{ij} \approx y_t^2  V^*_{3i} V_{3j}\,,  
\end{equation} 
where  a basis is used in which the  $d$-type quark Yukawa is diagonal.

There are two strict predictions in this general 
class of models  which have to be tested.  First, the MFV hypothesis implies 
the  usual CKM relations between $b \to s$, $b \to d$, 
and  $s \to d$ transitions. For example, this relation allows 
for upper bounds  on NP  effects in 
${\rm BR}(\bar B \to X_d\gamma)$, and ${\rm BR}(\bar B \to X_s \nu\bar \nu)$ using experimental data 
or bounds from ${\rm BR}(\bar B  \to X_s\gamma)$, and 
${\rm BR}(K \to \pi^+ \nu\bar \nu)$,  respectively. 
This emphasizes the need for 
high-precision measurements of $b \to  s/d$ , but also of 
$s \to  d$ transitions such as  the rare kaon decay 
$K \to \pi \nu\bar\nu$. 

The second prediction is  that  the CKM phase is the only   
source of CP violation. This implies that any phase 
measurement as  in 
$B \to \phi K_s$ or $\Delta M_{B_{(s/d)}}$ is  not sensitive 
to  new physics. 
This is an  additional assumption because the breaking of the flavour group and 
the discrete  CP symmetry is  in principle not connected at all. 
For example there is also a
renormalization-group invariant extension of the MFV concept allowing for flavour-blind phases 
as was shown in Ref.~\cite{Hurth:2003dk}; however these  lead to non-trivial CP  effects,  
which get  strongly constrained by flavour-diagonal observables 
such as   electric dipole moments~\cite{Hurth:2003dk}. 
So within the model-independent effective field theory approach of MFV we keep
the minimality  condition regarding CP. 
But in specific models like MSSM the discussion of additional CP phases within the MFV
framework makes sense and can also allow  for  a natural solution of the well-known 
supersymmetric CP
problem,  see for example Refs.~\cite{Mercolli:2009ns,Paradisi:2009ey}.

Scenarios with two Higgs doublets with large 
$\tan \beta = O(m_t/m_b)$  allow for  the unification of top and bottom 
Yukawa couplings as predicted in grand-unified models and 
for sizable new effects in helicity-suppressed decay models.
There are more general MFV relations existing  in this scenario due 
to the dominant role of scalar operators. However, since  
$\tan \beta$ is large,  
there is a new combination of spurions numerically relevant in 
the construction of higher-order MFV effective \mbox{operators,} namely

\begin{equation}
(Y_D Y_D^\dagger)_{ij} \approx y_d^2  \delta_{ij}\,,
\end{equation} 
which invalidates the general MFV relation between $b \to s/d$ and $s \to d$ transitions.

For more details we refer to the very recent complete mini-review on MFV~\cite{Isidori:2012ts}. Here we only add two issues on the application of the MFV hypothesis to the 
minimal supersymmetric standard  model (MSSM). Most interestingly, the MFV hypothesis can serve as a substitute for R-parity in the MSSM~\cite{Nikolidakis:2007fc,Csaki:2011ge}.
MFV  is sufficient to forbid a  too fast proton decay because when the MFV hypothesis is applied to R-parity violating terms, the spurion expansion leads to a suppression by 
neutrino masses and light-charged fermion masses, in this sense MFV within the MSSM  can be regarded a natural theory for R-parity violation. 
Secondly, the MFV framework is renormalization-group invariant by construction, however, it is not clear that the hierarchy  between the spurion terms is preserved when running down 
from the high scale to the low electroweak scale. Without this conservation of hierarchy, the MFV hypothesis would lose its practicability. However, as explicitly shown in 
Refs.~\cite{Paradisi:2008qh,Colangelo:2008qp}, a MFV-compatible change of the boundary conditions at the high scale has barely any influence on the low-scale spectrum.    
Finally, the MFV hypothesis solves the NP flavour problem only formally.   One still has to find explicit dynamical structures to realize the MFV hypothesis like gauge-mediated supersymmetric theories. And of course  the MFV hypothesis is not a theory of flavour; it does not explain the hierarchical structure
of the CKM matrix and the large mass splittings of the SM fermions.

\subsection{Effective  hamiltonian} 
Our analysis is based on the following MFV effective hamiltonian relevant to $b \to s$ transitions (and also for $b \to d$ transitions 
with obvious replacements)~\cite{D'Ambrosio:2002ex}:
\begin{eqnarray}
{\cal H}^{ b\to s}_{\rm eff} &=& -\frac{4 G_F}{\sqrt{2}}
 [  V^*_{us} V_{ub} (C^c_1 P^u_1 + C^c_2 P^u_2)
  + V^*_{cs} V_{cb} (C^c_1 P^c_1 + C^c_2 P^c_2)]
\nonumber \\
&-& \frac{4 G_F}{\sqrt{2}}  {  \sum_{i=3}^{10} [(V^*_{us} V_{ub}
+ V^*_{cs} V_{cb}) C^c_i \; + \; V^*_{ts} V_{tb} C^t_i] P_i +
V^*_{ts} V_{tb}
C^\ell_{0} P^\ell_{0}}~+~{\rm h.c.}
\label{eq:newHeff}
\end{eqnarray}
with
\begin{equation}
\begin{array}{ll}
P^u_1 =  (\bar{s}_L \gamma_{\mu} T^a u_L) (\bar{u}_L \gamma^{\mu}
T^a b_L)~, &
\vspace*{0.3cm}
P_5 =  (\bar{s}_L \gamma_{\mu_1}
                   \gamma_{\mu_2}
                   \gamma_{\mu_3}    b_L)\sum_q (\bar{q} \gamma^{\mu_1}
                                                         \gamma^{\mu_2}
                                                         \gamma^{\mu_3}     q)~,
 \\
P^u_2 =  (\bar{s}_L \gamma_{\mu}     u_L) (\bar{u}_L \gamma^{\mu}
b_L)~, & \vspace*{0.3cm}
P_6 =  (\bar{s}_L \gamma_{\mu_1}
                   \gamma_{\mu_2}
                   \gamma_{\mu_3} T^a b_L)\sum_q (\bar{q} \gamma^{\mu_1}
                                                          \gamma^{\mu_2}
                                                          \gamma^{\mu_3} T^a q)~,\\
P^c_1 =  (\bar{s}_L \gamma_{\mu} T^a c_L) (\bar{c}_L \gamma^{\mu}
T^a b_L)~,
& \vspace*{0.3cm}
P_7  =   \frac{e}{16\pi^2} m_b (\bar{s}_L \sigma^{\mu \nu}     b_R)
F_{\mu \nu}~,\\
P^c_2 =  (\bar{s}_L \gamma_{\mu}     c_L) (\bar{c}_L \gamma^{\mu}
b_L)~,
&\vspace*{0.3cm}
P_8  =   \frac{g_s}{16\pi^2} m_b (\bar{s}_L \sigma^{\mu \nu} T^a b_R)
G_{\mu \nu}^a~,\\
P_3 =  (\bar{s}_L \gamma_{\mu}     b_L) \sum_q (\bar{q}\gamma^{\mu}
q)~,
&\vspace*{0.3cm}
P_9  =   \frac{e^2}{16\pi^2} (\bar{s}_L \gamma_{\mu} b_L) \sum_\ell
 (\bar{\ell}\gamma^{\mu} \ell)~,\\
P_4 =  (\bar{s}_L \gamma_{\mu} T^a b_L) \sum_q (\bar{q}\gamma^{\mu}
T^a q)~,
&\vspace*{0.3cm}
P_{10} =  \frac{e^2}{16\pi^2} (\bar{s}_L \gamma_{\mu} b_L) \sum_\ell
                             (\bar{\ell} \gamma^{\mu} \gamma_5 \ell)~.\\
\end{array}
\end{equation}
In addition we have the following scalar-density operator with right-handed $b$-quark:\footnote{
Within the MFV framework, 
the corresponding Wilson coefficient $C_0^\ell$ is related to $C_{Q_1}=m_b C_S$ and $C_{Q_2}=m_b C_P$ by 
$C_0^\ell = 2\, C_{Q_1} = -2\, C_{Q_2}$,
where the operators are defined as $Q_1= Q_S/m_b =\frac{e^2}{(16 \pi^2)} (\bar{s}_L b_R)(\bar{\ell}\,\ell),\,\,\,
Q_2 = Q_P/m_b  =   \frac{e^2}{(16\pi^2)} (\bar{s}_L  b_R)(\bar{\ell}\gamma_5 \ell)$.}
\begin{equation}
P^\ell_{0} = \frac{e^2}{16\pi^2} (\bar s_L b_R) (\bar \ell_R \ell_L)~,
\end{equation}

There is no reason to update the MFV analysis of  precision $s \to d$ transitions of Ref.~\cite{Hurth:2008jc}.  
For completeness we state that for 
the rare decays $K\to \pi \nu\bar\nu$, we have 
the   following simple effective hamiltonian, 
\begin{equation}
\mathcal{H}^{s\to d}_{\rm eff}=\frac{G_{F}\alpha_{\rm em}\left( m_{Z}\right)
}{\sqrt{2}}{{\sum_{\ell=e,\mu,\tau}}}\left(  \frac{y_{\nu}}{2\pi\sin^{2}\theta_{W}}P_{\nu\bar{\nu}%
}\right)  +\mathrm{h.c.} \;,
\end{equation}
with the operator $P_{\nu\bar\nu}$ and the corresponding possible NP contribution $\delta C_{\nu\bar\nu}$,
\begin{eqnarray}
P_{\nu\bar{\nu}} =\left(  \bar{s}\gamma_{\mu}d\right)  \left(
\bar{\nu }_{\ell}\gamma^{\mu}\left(  1-\gamma_{5}\right)
\nu_{\ell}\right)\,,\,\, y_{\nu} ={\frac{1}{\vert V_{us}\vert}}\left(
\lambda_{t} (X_{t} + \delta C_{\nu\bar\nu})+{Re}\lambda_{c}\widetilde{P}_{u,c}\right)\;,
\label{NUCoeff}
\end{eqnarray}
and  $\lambda_{q}=V_{qs}^{\ast}V_{qd}$, 
$X_{t}=1.464\pm0.041$, $\widetilde{P}_{u,c}=(0.2248)^4\,P_{u,c}$~\cite{Antonelli:2008jg} and 
$P_{u,c}=0.41\pm0.04$~\cite{Brod:2008ss,Buras:2006gb,Mescia:2007kn}.

We follow here the analysis of  Ref.~\cite{Hurth:2008jc} and consider NP in the FCNC 
operators $P_7$,$P_8$,$P_9$,$P_{10}$ and in the two scalar operators $P_{\nu\bar\nu}$ and $P_0^\ell$ only;   
as argued, in principle most of the possible NP contributions to the   four-quark operators $P_{1-6}$  
could be reabsorbed into the Wilson coefficient of the FCNC operators.  The NP  contributions to the 
Wilson coefficients are parameterized as:
\begin{equation}
\delta C_i(\mu_b) = C_i^{\rm MFV}(\mu_b) - C_i^{\rm SM}(\mu_b)\;.
\end{equation}
where the $C_i^{\rm SM}(\mu_b)$ are given in Table~\ref{tab:wilson}.
\begin{table}
\begin{center}
\footnotesize{\begin{tabular}{|c|c|c|c|c|}\hline
$C_7^{\text{eff}}(\mu_b)$ & $C_8^{\text{eff}}(\mu_b)$ & $C_9(\mu_b)$ & $C_{10}(\mu_b)$ & $C_0^\ell(\mu_b)$  \\ \hline
-0.2974  & -0.1614  & 4.2297 & -4.2068 & 0  \\ \hline
\end{tabular}}
\caption{SM Wilson coefficients at $\mu_b=m_b^{\text{pole}}$ and $\mu_0=2M_W$ to NNLO accuracy in $\alpha_s$. \label{tab:wilson}}
\end{center}
\end{table}
%

\section{Observables and theoretical uncertainties}
\label{sec:obs}
We present the various $\Delta F =1$ observables which we use in our MFV fit or which we want to constrain or predict. 
We focus on their  dependence on the (non-standard) Wilson coefficients of the MFV effective theory and on the  main sources of the theoretical
uncertainty. 

\subsection{Radiative decay $\bar B \to X_{s,d} \gamma$}
The branching fraction for $B\to X_q \gamma$ ($q=s,d$) for a photon energy cut $E_\gamma > E_0$ can be
parameterized as
\begin{equation}
{\rm BR} (B \to X_q \gamma)_{E_\gamma > E_0}
=  {\rm BR}  (B \to X_c e \bar \nu)_{\rm exp} \, {6 \alpha_{\rm em} \over \pi C} \,
\left| V_{tq}^* V_{tb}^{}\over V_{cb}^{}\right|^2 
\, \Big[ P(E_0) + N(E_0) \Big] ,  \\
\label{br}
\end{equation}
where $\alpha_{\rm em} = \alpha_{\rm em}^{\rm on~shell}$~\cite{Czarnecki:1998tn}, $C = |V_{ub}|^2 / |V_{cb}|^2 \,\times\, \Gamma[B\to X_c e
  \bar \nu] / \Gamma [ B\to X_u e \bar \nu] $ and $P(E_0)$ and
$N(E_0)$ denote the perturbative and  nonperturbative contributions,
respectively. The latter are  normalized to the {\it charmless} semileptonic
rate  to separate the charm  dependence. 
The perturbative part of the branching ratio of $\bar B \to X_s \gamma$ is known to NNLL precision~\cite{Misiak:2006zs}, 
while the nonperturbative corrections are now estimated 
 to be well below $10\%$~\cite{Benzke:2010js}.
The overall uncertainty consists of nonperturbative (5\%), parametric
(3\%), perturbative (scale) (3\%) and $m_c$-interpolation ambiguity
(3\%), which are  added in quadrature.
An additional scheme dependence in the
determination of the pre-factor $C$ has been  found~\cite{Gambino:2008fj};  it is within the
perturbative uncertainty of $3\%$~\cite{Misiak:2008ss}. 
The dependence of the dominating perturbative part from the Wilson coefficients 
can be parametrized~\cite{Misiak:2006ab} at NNL: 
\begin{eqnarray}
P(E_0) = P^{(0)}(\mu_b) + \left(\frac{\alpha_s(\mu_b)}{4\pi}\right) \left[ P_1^{(1)}(\mu_b) + P_2^{(1)}(E_0,\mu_b) \right]  + {\cal O}\left(\alpha_s^2(\mu_b)\right),\,\, \mbox{\rm where}
\end{eqnarray}
\begin{eqnarray}
P^{(0)}(\mu_b) &=& \left[ C_7^{(0)\rm eff}(\mu_b)\right]^2, \,\,\,\,\,  P_1^{(1)}(\mu_b) = 2\, C_7^{(0)\rm eff}(\mu_b) \,C_7^{(1)\rm eff}(\mu_b)\;,\nonumber\\
P_2^{(1)}(E_0,\mu_b) &=& \sum_{i,j=1}^{8} C_i^{(0)\rm eff}(\mu_b)\; C_j^{(0)\rm eff}(\mu_b) \; K_{ij}^{(1)}(E_0,\mu_b).
\end{eqnarray}
The functions $K_{ij}^{(1)}$ can be found in Ref.~\cite{Misiak:2006ab}. The effective Wilson coefficients are given in the Appendix. 
We stress that we have  used NNLL precision (means inclusion of ${\cal O}\left(\alpha_s^2(\mu_b)\right)$ terms)  in our numerical analysis. 

The branching ratio of $\bar B \to X_d \gamma$ is only known to NLL QCD precision~\cite{Hurth:2003dk}. 
The error at this order is dominated by a large scale renormalization uncertainty of more than $12\%$
and by uncertainties due to CKM matrix elements of $10\%$. However, in view of the large experimental error the NLL
precision is still appropriate within our analysis.

\subsection{Isospin asymmetry $\Delta_0(B \to K^*\gamma)$}
Another  important observable which is already measured  is  the isospin breaking ratio.  
It arises when the photon is emitted from the spectator quark: 
\begin{equation}
\Delta_{0\pm}=\dfrac{\Gamma(\bar B^0\to\bar K^{*0}\gamma) - \Gamma(B^\pm \to K^{*\pm}\gamma)}{\Gamma(\bar B^0\to\bar K^{*0}\gamma) + \Gamma(B^\pm\to K^{*\pm}\gamma)}\;,
\end{equation}\\
where the partial decay rates are $CP$-averaged.  In the SM
spectator-dependent effects enter only at the order $\Lambda/m_b$,  whereas
isospin-breaking in the form factors is  expected to be a negligible
effect. Therefore, the SM prediction is as small as
{$O(8\%)$}.
Moreover, a part of the $\Lambda/m_b$ (leading) contribution cannot be calculated within the 
QCDf approach what leads to a large uncertainty~\cite{Kagan:2001zk}.  
However, the ratio is shown to be especially sensitive to NP
effects in the penguin sector.
The isospin asymmetry can be written as~\cite{Kagan:2001zk}:
\begin{equation}
\Delta_{0} ={\rm Re}(b_d-b_u) \;,
\end{equation} 
where the spectator dependent coefficients $b_q$ take the form:~\footnote{In this subsection we use the Wilson coefficients of  the traditional basis~\cite{Buchalla:1995vs}.}
\begin{equation}
b_q = \frac{12\pi^2 f_B\,Q_q}{\overline{m}_b\,T_1^{B\to K^*} a_7^c}\left(\frac{f_{K^*}^\perp}{\overline{m}_b}\,K_1+ \frac{f_{K^*} m_{K^*}}{6\lambda_B m_B}\,K_{2q}\right)\;.
\end{equation}
Here the coefficient $a_7^c$ reads \cite{Bosch:2001gv}:
\begin{eqnarray}
a^c_7(K^*\gamma) &=& C_7(\mu_b) + \frac{\alpha_s(\mu_b) C_F}{4\pi} \Big[ C_2(\mu_b) G_2(x_{cb})+ C_8(\mu_b) G_8\Big]\label{a7c}\\
&& +\frac{\alpha_s(\mu_h) C_F}{4\pi} \Big[ C_2(\mu_h) H_2(x_{cb})+ C_8(\mu_h) H_8\Big] \nonumber\;, 
\end{eqnarray}
where $\mu_h=\sqrt{\Lambda_h \mu_b}$ is the spectator scale. The functions $G_2, G_8, H_2,$ and $H_8$ can be found in Ref.~\cite{Bosch:2001gv}.
The functions $K_1$ and $K_{2q}$ can be written in function of the Wilson coefficients $C_i$  at scale $\mu_b$~\cite{Kagan:2001zk}:
\begin{eqnarray}
K_1 &=& -\left( C_6(\mu_b) + \frac{C_5(\mu_b)}{N} \right) F_\perp+ \frac{C_F}{N}\,\frac{\alpha_s(\mu_b)}{4\pi}\,\left\lbrace\left( \frac{m_b}{m_B} \right)^2 C_8(\mu_b)\,X_\perp \right.\\
&& \left. -C_2(\mu_b) \left[  \left(\frac43\ln\frac{m_b}{\mu_b} + \frac23 \right) F_\perp - G_\perp(x_{cb})\right] + r_1 \right\rbrace\nonumber\;,\\
\nonumber\\ 
K_{2q} &=& \frac{V_{us}^* V_{ub}}{V_{cs}^* V_{cb}}\left( C_2(\mu_b) + \frac{C_1(\mu_b)}{N} \right) \delta_{qu} + \left( C_4(\mu_b) + \frac{C_3(\mu_b)}{N} \right) \\
&&+ \frac{C_F}{N}\,\frac{\alpha_s(\mu_b)}{4\pi} \left[ C_2(\mu_b)\left( \frac43\ln\frac{m_b}{\mu_b} + \frac23 - H_\perp(x_{cb}) \right) + r_2 \right] \nonumber\;,
\end{eqnarray}
where $x_{cb}=\displaystyle\frac{m^2_c}{m^2_b}$ and $N=3$ and $C_F=4/3$ are colour factors. 
The convolution integrals of the hard-scattering kernels with the meson distribution amplitudes $F_\perp, G_\perp, H_\perp,$ and $X_\perp$ can be found in 
Ref.~\cite{Kagan:2001zk}, also the residual NLO corrections $r_1$ and $r_2$.

\subsection{Leptonic decays $B_{s,d} \to \mu^+\mu^-$}
The rare decay $B_s \to \mu^+ \mu^-$ proceeds via $Z^0$ penguin and box diagrams in the SM.  It is highly helicity-suppressed.  
However,  for large values of $\tan\beta$ this decay can receive large contributions.
In general, within MFV the pure leptonic decay $B_{s} \to \ell^+ \ell^-$
receive contributions only from the effective operators $P_{10}$ 
and $P^\ell_{0}$. These are free from the contamination of 
four-quark operators, which makes the generalization to the  
$b \to d$ case straightforward. 
The branching fraction in the MFV framework is given by 
\begin{eqnarray}
&&\hspace*{-0.7cm} \mathrm{BR}(B_s \to \mu^+ \mu^-) = \frac{G_F^2 \alpha^2}{64 \pi^3} f_{B_s}^2 \tau_{B_s} m_{B_s}^3 |V_{tb}V_{ts}^*|^2 \sqrt{1-\frac{4 m_\mu^2}{m_{B_s}^2}} \\
&&\hspace*{-0.5cm} \times \left[\left(1-\frac{4 m_\mu^2}{m_{B_s}^2}\right) \left| {\left(\frac{m_{B_s}}{m_b+m_s}\right)} (C_0^\mu/2) \right|^2 +  \left| {\left(\frac{m_{B_s}}{m_b+m_s}\right)}(C_0^\mu/(-2)) + 2 \, (C_{10}) \frac{m_\mu}{m_{B_s}}  \right|^2\right] \;, \nonumber
\end{eqnarray} 
where $f_{B_s}$ is the $B_s$ decay constant, $m_{B_s}$ is the $B_s$ meson mass and $\tau_{B_s}$ is the $B_s$ mean life.

The main theoretical uncertainty comes from the $B_s$ decay constant $f_{B_s}$,
which has recently been re-evaluated by independent lattice QCD groups
of Table~\ref{tab:lattice}.
\begin{table}
 \begin{center}
\begin{tabular}{|ll|l|l|}
 \hline
Lattice QCD Group & Ref.& $f_{B_s}$& $f_B$\\\hline
ETMC-11& \cite{Dimopoulos:2011gx}&$232\pm10$ MeV &$195\pm12$ MeV\\
Fermilab-MILC-11& \cite{Bazavov:2011aa,Neil:2011ku}&$242\pm9.5$ MeV &$197\pm9$ MeV\\
HPQCD-12&\cite{Na:2012kp}&$227\pm10$ MeV &$191\pm9$ MeV\\
\hline
Average && $234\pm10$ MeV &$194\pm10$ MeV\\
\hline
\end{tabular}
\caption{Average of lattice QCD results used in this work.  \label{tab:lattice}}
\end{center}
\end{table}
Their 4.3\% uncertainties agree, as do their results within these
uncertainties, so that {following Ref.~\cite{Mahmoudi:2012un}} we have chosen an average of these three results
in what follows. This implies a 8.7\% uncertainty on the
branching ratio. 
The most important parametric uncertainty comes from  the CKM matrix element 
$V_{ts}$ with  $5\%$.

Within the MFV scenario the $B_{d} \to \ell^+ \ell^-$ rate can be obtained  
from the one of   $B_{s} \to \ell^+ \ell^-$  with the exchange  
($V_{ts}$, $m_{B_s}$, $m_s${, $f_{B_s}$})  $ \to$ ($V_{td}$, $m_{B_d}$, $m_d${, $f_{B_d}$}).
This implies a very important MFV relation ({$O(m_d/m_s)$} are neglected), 
\begin{equation}
\frac{ \Gamma(B_s\to \ell^+ \ell^-) }{ \Gamma(B_d\to \ell^+ \ell^-) }
\approx \frac{  f_{B_s}m_{B_s} }{ f_{B_d}m_{B_d} }
\left| \frac{ V_{ts} }{ V_{t_d} } \right|^2~.
\label{bllcorrelation}
\end{equation}

\subsection{Inclusive $\bar B \to X_s \mu^+\mu^-$ and $\bar B \to X_s \tau^+\tau^-$}
The decay $B \rightarrow X_s \ell^+\ell^-$ is particularly attractive
because it offers several kinematic observables.  The angular
decomposition of the decay rate provides three independent observables,
$H_T$, $H_A$ and $H_L$, from which one can extract the short-distance
electroweak Wilson coefficients that test for
NP:
\begin{equation}\label{eq:d3Gamma}
\frac{d^3\Gamma}{d q^2\,  d z}
= \frac{3}{8} \Bigl[(1 + z^2) H_T(q^2)
+ 2(1 - z^2) H_L(q^2)
+  2 z H_A(q^2)
\Bigr]
\,.\end{equation}
Here $z=\cos\theta_\ell$, $\theta_\ell$ is the angle between the negatively
charged lepton and the $\bar B$ meson in the center-of-mass frame of
the dilepton system, and $q^2$ is  the dilepton mass squared.  $H_A$ is equivalent
to the forward-backward asymmetry, and the dilepton-mass spectrum is
given by $H_T + H_L$.  The observables mainly constrain   the Wilson coefficients
$C^{\rm eff}_7$, $C^{\rm eff}_9$ and $C^{\rm eff} _{10}$. 

One defines perturbatively dominated (means theoretically clean) observables 
within two dilepton-mass windows avoiding the region with the $c\bar c$ resonances:
the low-$q^2$ region (1 GeV$^2  < q^2 < 6$ GeV$^2$) and  the high-$q^2$ region  ($q^2 > 14.4$ GeV$^2 $).  

In order to show the dependence of the observables on the Wilson coefficients, we use the 
conventions of Ref.~\cite{Ghinculov:2003qd}. We note that for the numerical evaluation we have used the conventions in Ref.~\cite{Huber:2007vv}, 
in particular we have chosen   the charmless semileptonic decay rate as normalization.  
For the branching ratio within the MFV framework we find:
\begin{eqnarray}\nonumber \label{branching}
&&\frac{d{\rm BR}(B\rightarrow X_s\ell^+\ell^-)}{d\hat{s}}={\rm BR}(B \rightarrow X_c\ell\bar{\nu})\frac{\alpha^2}{4\pi^2f(z)\kappa(z)}\frac{|V_{tb}V_{ts}^*|^2}{|V_{cb}|^2}(1-\hat{s})^2
\sqrt{1-\frac{4\hat{m}^2_{\ell}}{\hat{s}}}\\\nonumber
&&\times\Biggl\lbrace|C_9^{new}|^2(1+\frac{2\hat{m}^2_{\ell}}{\hat{s}})(1+2\hat{s})\left(1+\frac{\alpha_s}{\pi}\tau_{99}(\hat{s})\right)+4|C_7^{new}|^2 
(1+\frac{2\hat{m}^2_{\ell}}{\hat{s}})(1+\frac{2}{\hat{s}})\left(1+\frac{\alpha_s}{\pi}\tau_{77}(\hat{s})\right)\\\nonumber
&+&|C_{10}^{new}|^2[(1+2\hat{s})+\frac{2\hat{m}^2_{\ell}}{\hat{s}}(1-4\hat{s})]\left(1+\frac{\alpha_s}{\pi}\tau_{99}(\hat{s})\right)+12\mathrm{Re}(C_7^{new}C_9^{new*})(1+\frac{2\hat{m}^2_{\ell}}{\hat{s}})\left(1+\frac{\alpha_s}{\pi}\tau_{79}(\hat{s})\right)\\
&+&\frac{3}{4}|C_{0}^\ell|^2(\hat{s}-2\hat{m}^2_{\ell})
- 3 Re(C_{10}^{new}C_{0}^{\ell *})\hat{m_{\ell}}\Biggr\rbrace\label{BR_BXsll}
+ \delta^{brems}_{d\mathcal{B}/d\hat{s}} + \delta^{1/m_b^2}_{d\mathcal{B}/d\hat{s}}+ \delta^{1/m_b^3}_{d\mathcal{B}/d\hat{s}} +\delta^{1/m_c^2}_{d\mathcal{B}/d\hat{s}}
+\delta^{em}_{d\mathcal{B}/d\hat{s}}\;,\\\nonumber
\end{eqnarray}
where  the hat  indicates a normalization by $m_b$. The functions $\tau_i$ correspond to specific bremsstrahlung terms. As  indicated,  further (but finite) bremsstrahlung, electromagnetic
and power corrections have to be  added, see Ref.~\cite{Ghinculov:2003qd}  for more details. Our formula~(\ref{branching}) is  consistent with the results in Ref.~\cite{Grossman:1996qj}.

For the dependence of the forward-backward asymmetry on the Wilson coefficients we find within the MFV setting: 
\begin{eqnarray}\nonumber\label{AFB}
A_{FB}(\hat{s})&=&\int_0^1dz\frac{d^2{\rm BR}}{d\hat{s}dz}-\int_{-1}^0dz
\frac{d^2{\rm BR}}{d\hat{s}dz}
=-{\rm B}(B\rightarrow X_c\ell\bar{\nu})\frac{3\alpha^2}{4\pi^2f(z)\kappa(z)}\frac{|V_{tb}V_{ts}^*|^2}{|V_{cb}|^2} (1-\hat{s})^2(1-\frac{4\hat{m}^2_{\ell}}{\hat{s}})\\
&\times&\Biggl\lbrace\mathrm{Re}(C_9^{new}C_{10}^{new*})\hat{s}\left(1+\frac{\alpha_s}{\pi}\tau_{910}(\hat{s})\right)+2\mathrm{Re}(C_7^{new}C_{10}^{new*})\left(1+\frac{\alpha_s}{\pi}\tau_{710}(\hat{s})\right)\\\nonumber
&+&\mathrm{Re}(\,(C_9^{new}/2 + C_7^{new})\, C_{0}^{\ell *})\hat{m}_{\ell}\Biggr\rbrace +\delta^{1/m_b^2}_{A_{FB}}(\hat{s}) +\delta^{1/m_c^2}_{A_{FB}}(\hat{s}) +\delta^{brems}_{A_{FB}}(\hat{s})+ \delta^{em}_{A_{FB}}(\hat{s})\;.\\
\nonumber
\end{eqnarray}
We note that Eq.~(\ref{AFB}) is consistent with the results in Ref.~\cite{Grossman:1996qj}, but disagrees  with Refs.~\cite{Xiong:2000cp,Xiong:2001up}
for the scalar contributions.
The {\it new} Wilson coefficients are defined in Ref.~\cite{Ghinculov:2003qd} and are given in the Appendix.

In the low-$q^2$ region the theoretical uncertainty is around $7\%$ for the branching ratio, however there is an 
additional $5\%$ uncertainty due to nonlocal power corrections to be added~\cite{Huber:2007vv}. 
In the high-$q^2$  region, one encounters the breakdown of the 
heavy-mass expansion at the endpoint. However, for an integrated high-$q^2$ spectrum an effective expansion exists 
in inverse powers of $m_b^{\rm eff} = m_b \times (1 - \sqrt{\s_{\rm min}})$ rather than $m_b$.
The resulting large  theoretical uncertainties in the high-$q^2$ due to the power corrections of around $25\%$ could be 
significantly reduced by normalizing
the $\bar B \rightarrow X_s \ell^+ \ell^-$ decay rate to the
semileptonic $\bar B \rightarrow X_u \ell \bar\nu$ decay rate with the
same $q^2$ cut~\cite{Ligeti:2007sn}.
For example,   the uncertainty due to the dominating $1/m_b^3$ term
would be   reduced from $19\%$ to $9\%$~\cite{Huber:2007vv}.

\subsection{Exclusive decay $B \to K^* \ell\ell$}
The exclusive semi-leptonic penguin modes offer a larger variety of experimentally accessible observables 
than do the inclusive ones, but the hadronic uncertainties in the theoretical predictions are in general 
larger.  

The physics opportunities of $B \to K^{*}\ell\ell$ ($\ell=e, \mu,
\tau$) decays depend strongly on the measurement of their angular distributions.
This decay with $K^*$ on the mass shell
has a 4-fold differential distribution~\cite{Kruger:1999xa, Altmannshofer:2008dz}
\begin{equation}
  \label{eq:differential decay rate}
  \frac{d^4\Gamma[B \to K^{*}(\to K \pi)\ell\ell]}
       {d q^2\, d\ctl\, d\ctk\, d\phi} =
  \frac{9}{32\pi} \sum_i J_i(q^2)\, g_i(\theta_l, \theta_K, \phi)\,,
\end{equation}
w.r.t. the dilepton invariant mass $q^2$ and the angles $\theta_l$, $\theta_K$, 
and $\phi$ (as defined in~\cite{Egede:2008uy}).
It offers 12  
observables $J_i(q^2)$, from which all other known ones 
can be derived upon integration over appropriate combinations of angles. 

The $J_i$ depend on products of the eight  theoretical complex $K^*$ spin amplitudes 
$A_i$, $A_{\bot,\|,0}^{L,R},A_t,A_S$.  The $J_i$ are bi-linear 
functions of the spin amplitudes such as
\begin{equation}
  \label{eq:J1s}
  J_{s}^1 = \frac{3}{4} \left[|\apeL|^2 + |\apaL|^2 + |\apeR|^2 + |\apaR|^2  \right],
\end{equation}
with the expression for the eleven other $J_i$ terms given for example 
in~\cite{Kruger:2005ep,Altmannshofer:2008dz,Egede:2010zc}.

The dilepton invariant mass spectrum for $B \to {K}^* \ell^+ \ell^-$ can be recovered after integrating the 4-differential distribution
over all angles, while the (normalized) forward-backward asymmetry $A_{\rm FB}$ can be 
defined after full $\phi $ and $\theta_{K^*}$ integration \cite{Bobeth:2008ij} ($J_i \equiv 2 J^s_i + J^c_i$):
\begin{equation}
\frac{d\Gamma}{dq^2} = \frac{3}{4} \bigg( J_1 - \frac{J_2}{3} \bigg),\;\,\,\,\, A_{\rm FB}(q^2) \equiv
     \left[\int_{-1}^0 - \int_{0}^1 \right] d\cos\theta_l\, 
          \frac{d^2\Gamma}{dq^2 \, d\cos\theta_l} \Bigg/\frac{d\Gamma}{dq^2}
 = -\frac{3}{8} \frac{J_6}{d\Gamma / dq^2}\;. 
\end{equation}
Moreover, the fraction of the longitudinal polarized $K^*$ is given by  $F_L = {(3\,J^c_1-J^c_2)}/{(4\,d\Gamma / dq^2)}$.
These  three observables represent the {\it early}  ones, which have been measured already by the $B$ factories and
now with much better precision  by the LHCb experiment. 

With more luminosity, theoretically  much cleaner angular observables will be available.  In the low- and high-$q^2$   region it is always appropriate to design 
optimized observables by using specifically chosen normalizations for the independent
set of observables.
In the low-$q^2$region, specific ratios of observables allow
for a complete cancellation of the hadronic uncertainties due to the  form factors
in leading order and, thus,  for a high increase in the sensitivity to new physics
structures~\cite{Egede:2008uy, Egede:2010zc}, for example
the transversity amplitudes: 
\begin{equation}   \label{eq:AT4:def}
  A_T^{(2)} =
 \frac{1}{2} \frac{J_3}{J_2^s} , \,\,\, 
  A_T^{(3)} = 
  \sqrt{\frac{4 J_4^2 + \beta_l^2 J_7^2}{- 2 J_2^c (2 J_2^s + J_3)}} ,\,\,\,
A_T^{(4)} = 
 \sqrt{ \frac{\beta_l^2 J_5^2 + 4 J_8^2}{4 J_4^2 + \beta_l^2 J_7^2}} .
\end{equation}
In the high-\qsq region, two groups of 
ratios of observables  can be constructed which dominantly depend either on 
short- or on long-distance physics~\cite{Bobeth:2010wg, Bobeth:2011gi}.
In addition to the $A_T^{(i)}$ observables some new transversity observables were proposed:
\begin{equation}
  H_T^{(1)}  =  
   \frac{\sqrt{2} J_4}{\sqrt{- J_2^c \left(2 J_2^s - J_3\right)}} ,\,\,\,
  \label{eq:def:HT1}
  H_T^{(2)}  = 
  \frac{\beta_l J_5}{\sqrt{-2 J_2^c \left(2 J_2^s + J_3\right)}} ,\,\,\,
  H_T^{(3)} =
   \frac{\beta_l J_6}{2 \sqrt{(2 J_2^s)^2 - J_3^2}}.
 \end{equation} 
In the high-$q^2$ region  $H_T^{(2,3)}$ depend only on short-distance information in leading order, while $F_L$ and 
$A_T^{(2,3)}$ depend only on long-distance quantities.~\footnote{There are three observables which 
are already measured beyond the early observables 
mentioned above: $S_3 = (J_3 + \bar J_3)/[d(\Gamma +\bar \Gamma)/d q^2]$, $A_{im}$,
and the isospin asymmetry, but all three observables have no significant impact on the MFV scenario yet.}

The theoretical treatment in the low- and high-\qsq is based on different theoretical
concepts. Thus, the consistency of the consequences out of the two sets of measurements
will allow for an important crosscheck. 

In the low-$q^2$ region, the up-to-date description of exclusive heavy-to-light $B \to K^* \ell^+\ell^-$ 
decays  is the method of QCD-improved Factorization (QCDF) and 
its field-theoretical formulation of Soft-Collinear Effective Theory 
(SCET). In the combined limit of a heavy $b$-quark and of an energetic $K^*$ meson,  the decay amplitude
factorizes to leading order in $\Lambda/m_b$ and to all orders in $\alpha_s$
into process-independent non-perturbative quantities like $B\to K^*$ form factors  
and light-cone distribution amplitudes (LCDAs) of the heavy (light) 
mesons and perturbatively calculable quantities, which are known to
$O(\alpha_s^1)$~\cite{Beneke:2001at,Beneke:2004dp}. 
Further, the {\it seven} 
a priori independent $B\to K^*$ QCD form factors reduce 
to {\it two}  universal {\it soft} form factors $\xi_{\bot,\|}$~\cite{Charles:1998dr}. 
The factorization formula applies well in the range
of the dilepton mass range, $1\; {\rm GeV}^2 < q^2 < 6\; {\rm GeV}^2$.

Taking into account all these  simplifications the various  \Kstar spin amplitudes at leading order in $\lqcd/m_b$ 
and \as turn out to be linear in the soft form factors $\xi_{\bot,\|}$ and also in the short-distance
Wilson coefficients which allows to design a set of optimized observables in which any soft form factor 
dependence (and its corresponding uncertainty)  cancels out for all low dilepton masses~\qsq at leading order in 
\as and $\lqcd/m_b$~\cite{Egede:2008uy, Egede:2010zc}:
\begin{subequations}
  \begin{align}
\apeLR &=\sqrt{2} N m_B(1- \sh)\bigg[  
(\Ceff9  \mp \C{10} )
+\frac{2\hat{m}_b}{\sh} \Ceff7  
\bigg]\xi_{\bot}(E_\kstar),   \\
\apaLR &= -\sqrt{2} N m_B (1-\sh)\bigg[
 (\Ceff9  \mp \C{10} ) 
+\frac{2\hat{m}_b}{\sh}  \Ceff7  \bigg] \xi_{\bot}(E_\kstar)\, , \\
\azeLR  &= -\frac{Nm_B }{2 \hat{m}_\kstar \sqrt{\sh}} (1-\sh)^2\bigg[ (\Ceff9   \mp \C{10} ) 
+ 2
\hat{m}_b \Ceff7  \bigg]\xi_{\|}(E_\kstar)\, ,\\
A_t  &= \frac{Nm_B }{ \hat{m}_\kstar \sqrt{\sh}} (1-\sh)^2\bigg[ \C{10} -  \frac{q^2}{4 m_\ell{m_b}}  C_0^\ell   \bigg] \xi_{\|}(E_\kstar) \,,\\
A_S &=   \frac{N {m_B^2} }{ 2 \hat{m}_\kstar {m_b}}  (1-\sh)^2\bigg[(-1) C_0^\ell   \bigg] \xi_{\|}(E_\kstar) \,,
  \end{align}
\end{subequations}
with $\sh = \qsq/m_B^2$, $\hat{m}_i = m_i/m_B$. Here we neglect
terms of $O(\hat{m}_{K^*}^2)$ but we include these terms in our numerical analysis.   
The factor $N$ collects all pre-factors and can be found in Ref.~\cite{Egede:2008uy, Egede:2010zc}. 
The soft form factors are fixed in a specific factorization scheme using QCD sum rule techniques as discussed in the Appendix.  

However, in the {early}  observables, namely $d\Gamma/dq^2, A_{\rm FB}, F_{\rm L}$ there is still  
a large  theoretical uncertainty due to the form factors which do not cancel out to first order in these cases.

Within the QCDF/SCET approach, a general, quantitative method to estimate the
important $\lqcd/m_b$ corrections to the heavy quark limit is missing. 
In semileptonic decays a simple dimensional estimate of $10\%$  is often used. Under 
the assumption that the main part of the $\lqcd/m_b$ corrections is included in 
the full form factors, the difference of the theoretical
results using the full QCD form factors on one hand and the soft form factors on 
the other hand confirms this simple dimensional estimate. In fact, the comparison 
of the approaches leads to a $7\%$ shift of the central value.

The low-hadronic recoil region  is characterized by large values of the dilepton
invariant mass $\qsq \gsim (14 - 15) \gev^2$ above the two narrow resonances of
$\jpsi$ and $\psitwos$. It is shown that local operator product expansion
is applicable ($\qsq \sim m_b^2$)
\cite{Grinstein:2004vb, Beylich:2011aq} and it allows to obtain the $B \to K^* \ell^+\ell^-$
matrix element in  a systematic expansion in $\alpha_s$ and in $\Lambda/m_b$. 
Most important, the leading power corrections are shown to be suppressed 
by $(\lqcd/m_b)^2$ or $\as \lqcd/m_b$ \cite{Beylich:2011aq} and  to contribute 
only at the few percent level. 
The only caveat is that heavy-to-light form factors are known
only from  extrapolations from LCSR calculations at low-\qsq at present. But this may improve in the future 
when direct lattice calculations in the high-$q^2$ are available~\cite{Bobeth:2010wg}.

There are improved Isgur-Wise relations between the form factors in leading power
of $\Lambda/m_b$.  Their application and the introduction of specific modified Wilson coefficients
lead to simple expressions for the $K^*$ spin amplitudes to leading order in $1/m_b$ in the low
recoil region, for example we have~\cite{Bobeth:2010wg}
\begin{align}
  \label{eq:Aperp}
  A_\perp^{L,R} & = 
  +i \{(\wilson[eff,mod]{9} \mp \wilson{10}) + 
      \kappa\frac{2 \hat{m}_b}{\hat{s}}\, \wilson[eff,mod]{7}\}
      f_\perp ,
\\[1mm]
  A_\parallel^{L,R} & =
    -i \{(\wilson[eff,mod]{9} \mp \wilson{10}) + 
      \kappa\frac{2 \hat{m}_b}{\hat{s}}\, \wilson[eff,mod]{7}\} f_\parallel,
      \\
  \label{eq:A0}
  A_0^{L,R} & =
  -i \{\(\wilson[eff,mod]{9} \mp \wilson{10}\) + 
    \kappa \frac{2 \hat{m}_b}{\hat{s}}\, \wilson[eff,mod]{7}\} f_0 ,
    \end{align}
where the form factors $f_\perp, f_\parallel,$ and $f_0$ are linearly connected to the QCD form factors
(see  Ref.~\cite{Bobeth:2010wg}). 
The {\it modified} effective Wilson coefficients introduced in Ref.~\cite{Grinstein:2004vb} are given in the Appendix.
Then, the three considered observables at  leading order can be written in the high-$q^2$ region as~\cite{Bobeth:2010wg}
\begin{equation} \label{eq:dG:HQET}
  \frac{\dd \Gamma}{\dd q^2}  = 
  2\, \rho_1 \times (f_0^2 + f_\perp^2 + f_\parallel^2),\,\,\,
  A_{\rm FB}  =
  3\, \frac{\rho_2}{\rho_1} \times \frac{f_\perp f_\parallel}
  {(f_0^2 + f_\perp^2 + f_\parallel^2)},\,\,\,\
  F_{\rm L} = 
  \frac{f_0^2}{f_0^2 + f_\perp^2 + f_\parallel^2},
\end{equation}
where only the two independent combinations of Wilson coefficients enter, namely 
\begin{equation} 
  \rho_1  \equiv 
  \left|\wilson[eff]{9} + \kappa \frac{2\hat{m}_b}{\hat{s}}\wilson[eff]{7}\right|^2 
   + \left|\wilson{10}\right|^2 , \,\,\, 
  \rho_2  \equiv 
  \Re{\(\wilson[eff]{9} + \kappa \frac{2\hat{m}_b}{\hat{s}} \wilson[eff]{7}\) \wilson[*]{10}} .  
\end{equation}
$\rho_1$ and $\rho_2$ are shown to be largely $\mu$-scale independent~\cite{Grinstein:2004vb}.

As mentioned above, the leading power corrections
of the OPE arise at ${\cal{O}}(\alpha_s
\Lambda/m_b, m_c^4/Q^4)$ and of the order of a few percent.  The $\Lambda/m_b$ corrections to
the amplitudes from the form factor relations  are parametrically suppressed as
well, by small dipole coefficients, such that one  can estimate the leading power
correction from the form factor relations to the decay amplitudes as order $( 2
\wilson[eff]{7}/ \wilson[eff]{9} ) \Lambda/m_b$. So in general, the dominant
power corrections to the transversity amplitudes are of the order of a few percent
\cite{Bobeth:2010wg}.

\section{Numerical details}  

Within  our numerical analysis we use the most recent LHCb results for the exclusive decays $B_s\to\mu^+\mu^-$ and $B \to K^* \mu^+ \mu^-$, and Belle, Babar and CDF results for the other decays. The experimental values are provided in Table~\ref{tab:obs}. For comparison, we also consider the pre-LHCb data as given in Table~\ref{tab:obs-preLHCb}.

There is a remark in order regarding the branching fraction of $B_s\to\mu^+\mu^-$.  Its  value provided by the experiments corresponds to an  untagged value, while the theoretical predictions are CP-averaged. As pointed out recently in \cite{deBruyn:2012wj,deBruyn:2012wk}, the untagged branching ratio is related to the CP-averaged one by:
\begin{equation}
\mathrm{BR^{untag}}(B_s \to \mu^+ \mu^-)=\left[\frac{1+ \mathcal{A}_{\Delta\Gamma}\,y_s}{1-y_s^2}\right] \mathrm{BR}(B_s \to \mu^+ \mu^-)\;,
\end{equation}
where
\begin{equation}
y_s \equiv \frac12 \tau_{B_s} \Delta\Gamma_s = 0.088 \pm 0.014\;,
\end{equation}
and
\begin{equation}
\mathcal{A}_{\Delta\Gamma} = \frac{|P|^2 \cos(2\varphi_P) - |S|^2 \cos(2\varphi_S)}{|P|^2 + |S|^2}\;,
\end{equation}
with
\begin{equation}
S \equiv \sqrt{1-4 \frac{m_\mu^2}{M_{B_s}^2}} \, \frac{M_{B_s}^2}{2m_\mu} \, \frac{1}{m_b + m_s} \frac{C_0^\ell/2}{C_{10}^{SM}}\;,
\end{equation}
\begin{equation}
P \equiv \frac{C_{10}}{C_{10}^{SM}} + 
\frac{M_{B_s}^2}{2m_\mu} \, \frac{1}{m_b + m_s} \frac{-C_0^\ell/2}{C_{10}^{SM}}\;,
\end{equation}
and
\begin{equation}
\varphi_S = \arg(S)\;,\qquad \varphi_P = \arg(P)\;.
\end{equation}
The obtained branching ratio can then be directly compared to the experimental result. 
\begin{table}
\begin{center}
\footnotesize{\begin{tabular}{|l|l|l|}\hline
Observable & Experiment & SM prediction \\ \hline
BR($B \to X_s \gamma$) & $(3.55 \pm 0.24\pm 0.09)\times 10^{-4}$ \cite{Asner:2010qj} & $(3.08 \pm 0.24)\times 10^{-4}$\\ \hline
$\Delta_0(B \to K^* \gamma)$ & $(5.2 \pm 2.6\pm 0.09)\times 10^{-2}$ \cite{Asner:2010qj}  & $(8.0 \pm 3.9)\times 10^{-2}$\\ \hline
BR($B \to X_d \gamma$) & $(1.41 \pm 0.57)\times 10^{-5}$ \cite{delAmoSanchez:2010ae,Wang:2011sn} & $(1.49 \pm 0.30)\times 10^{-5}$ \\ \hline
BR($B_s \to \mu^+\mu^-$) & $< 4.5 \times 10^{-9}$ \cite{Aaij:2012ac} & $(3.53 \pm 0.38) \times 10^{-9}$\\ \hline
$\langle dBR/dq^2(B \to K^* \mu^+ \mu^-) \rangle_{q^2\in[1,6] \rm{GeV}^2}$ & $(0.42 \pm 0.04 \pm 0.04)\times 10^{-7}$ \cite{LHCb-CONF-2012-008} & $(0.47 \pm 0.27)\times 10^{-7}$\\ \hline
$\langle dBR/dq^2(B \to K^* \mu^+\mu^-) \rangle_{q^2\in[14.18,16] \rm{GeV}^2}$ & $(0.59 \pm 0.07 \pm 0.04)\times 10^{-7}$ \cite{LHCb-CONF-2012-008} & $(0.71 \pm 0.18)\times 10^{-7}$\\ \hline
$\langle A_{FB}(B \to K^* \mu^+ \mu^-) \rangle_{q^2\in[1,6] \rm{GeV}^2}$ & $-0.18 \pm 0.06 \pm 0.02$ \cite{LHCb-CONF-2012-008} & $-0.06 \pm 0.05$\\ \hline
$\langle A_{FB}(B \to K^* \mu^+\mu^-) \rangle_{q^2\in[14.18,16] \rm{GeV}^2}$ & $0.49 \pm 0.06 \pm 0.05$ \cite{LHCb-CONF-2012-008} & $0.44 \pm 0.10$\\ \hline
$q_0^2 (A_{FB}(B \to K^* \mu^+ \mu^-))$ & $4.9 ^{+1.1}_{-1.3} \;\mathrm{ GeV}^2$ \cite{LHCb-CONF-2012-008} & $4.26 \pm 0.34\; \mathrm{ GeV}^2$\\ \hline
$\langle F_{L}(B \to K^* \mu^+ \mu^-) \rangle_{q^2\in[1,6] \rm{GeV}^2}$ & $0.66 \pm 0.06 \pm 0.04$ \cite{LHCb-CONF-2012-008} & $0.72 \pm 0.13$\\ \hline
BR($B \to X_s \mu^+\mu^-)_{q^2\in[1,6] \rm{GeV}^2}$ & $(1.60 \pm 0.68)\times 10^{-6}$ \cite{Aubert:2004it,Iwasaki:2005sy} & $(1.78 \pm 0.16)\times 10^{-6}$\\ \hline
BR($B \to X_s \mu^+\mu^-)_{q^2>14.4 \rm{GeV}^2}$ & $(4.18 \pm 1.35)\times 10^{-7}$ \cite{Aubert:2004it,Iwasaki:2005sy} &  $(2.19 \pm 0.44)\times 10^{-7}$\\ \hline
\end{tabular}}
\caption{Input observables:  The experimental data represent the most recent one.  The updated SM predictions are based on the input parameters given in Table~\ref{tab:input}. \label{tab:obs}}
\label{Inputobservables}
\end{center}
\end{table}
\begin{table}
\begin{center}
\footnotesize{\begin{tabular}{|l|l|}\hline
Observable & Experiment \\ \hline
BR($B \to X_s \gamma$) & $(3.55 \pm 0.24\pm 0.09)\times 10^{-4}$ \cite{Asner:2010qj} \\ \hline
$\Delta_0(B \to K^* \gamma)$ & $(5.2 \pm 2.6\pm 0.09)\times 10^{-2}$ \cite{Asner:2010qj}  \\ \hline
BR($B \to X_d \gamma$) & $(1.41 \pm 0.57)\times 10^{-5}$ \cite{delAmoSanchez:2010ae,Wang:2011sn} \\ \hline
BR($B_s \to \mu^+\mu^-$) & $< 5.8 \times 10^{-8}$ \cite{Aaltonen:2007ad} \\ \hline
$\langle dBR/dq^2(B \to K^* \ell^+ \ell^-) \rangle_{q^2\in[1,6] \rm{GeV}^2}$ & $(0.32 \pm 0.11 \pm 0.03)\times 10^{-7}$ \cite{CDF} \\ \hline
$\langle dBR/dq^2(B \to K^* \mu^+\mu^-) \rangle_{q^2\in[14.18,16] \rm{GeV}^2}$ & $(0.83 \pm 0.20 \pm 0.07)\times 10^{-7}$ \cite{CDF} \\ \hline
$\langle A_{FB}(B \to K^* \mu^+ \mu^-) \rangle_{q^2\in[1,6] \rm{GeV}^2}$ & $ 0.43 \pm 0.36 \pm 0.06$ \cite{CDF} \\ \hline
$\langle A_{FB}(B \to K^* \mu^+\mu^-) \rangle_{q^2\in[14.18,16] \rm{GeV}^2}$ & $0.42 \pm 0.16 \pm 0.09$ \cite{CDF} \\ \hline
$\langle F_{L}(B \to K^* \mu^+ \mu^-) \rangle_{q^2\in[1,6] \rm{GeV}^2}$ & $0.50 \pm 0.30 \pm 0.03$ \cite{CDF} \\ \hline
BR($B \to X_s \mu^+\mu^-)_{q^2\in[1,6] \rm{GeV}^2}$ & $(1.60 \pm 0.68)\times 10^{-6}$ \cite{Aubert:2004it,Iwasaki:2005sy} \\ \hline
BR($B \to X_s \mu^+\mu^-)_{q^2>14.4 \rm{GeV}^2}$ & $(4.18 \pm 1.35)\times 10^{-7}$ \cite{Aubert:2004it,Iwasaki:2005sy} \\ \hline
\end{tabular}}
\caption{Experimental data used for pre-LHCb fit. \label{tab:obs-preLHCb}}
\end{center}
\end{table}

The SM predictions entering the MFV-fit are based on the theoretical analyses given in Section~\ref{sec:obs}. We have used 
the input parameters of Table~\ref{tab:input} and the program    {\tt SuperIso v3.3}~\cite{Mahmoudi:2007vz,Mahmoudi:2008tp}
in order to update SM predictions. They are given in Table~\ref{tab:obs}.

To obtain constraints on the Wilson coefficients, we scan over $\delta C_7$, $\delta C_8$, $\delta C_9$, $\delta C_{10}$ and $\delta C_0^\ell$. For each point, we then compute the flavour observables using {\tt SuperIso v3.3}~\cite{Mahmoudi:2007vz,Mahmoudi:2008tp} and compare with the experimental results by calculating $\chi^2$ as:
\begin{equation}
\chi^2 = \sum_i \frac{(O_i^{\rm exp} - O_i^{\rm th})^2}{(\sigma_i^{\rm exp})^2 + (\sigma_i^{\rm th})^2} \;,
\end{equation}
where $O_i^{\rm exp}$ and $O_i^{\rm th}$ are  the central values of the experimental result and theoretical prediction of observable $i$ respectively, and $\sigma_i^{\rm exp}$ and $\sigma_i^{\rm th}$ are the experimental and theoretical errors respectively. The global fits are obtained by minimization of the $\chi^2$. For LHCb fits, we consider the measurements of the observables given in Table~\ref{tab:obs} while for the pre-LHCb  the measurements of Table~\ref{tab:obs-preLHCb} are considered.
\begin{table}
\begin{center}
\footnotesize{\begin{tabular}{|lr|lr|}\hline
$m_B=5.27950$ GeV                         & \cite{Nakamura:2010}     &        $m_{B_s} = 5.3663 $ GeV& \cite{Nakamura:2010}                            \\
$m_{K^*}=0.89594$ GeV                     & \cite{Nakamura:2010}     & $|V_{tb}V_{ts}^*|=0.0403 ^{+0.0011}_{-0.0007}$         & \cite{Nakamura:2010}          \\ \hline
$m_b^{\overline{MS}}(m_b)=4.19 ^{+0.18}_{-0.06}$ GeV & \cite{Nakamura:2010}     & $m_c^{\overline{MS}}(m_c)=1.29 ^{+0.05}_{-0.11}$ GeV   & \cite{Nakamura:2010}\\ 
$m_t^{pole}=172.9 \pm0.6 \pm0.9$ GeV       & \cite{Nakamura:2010}     &$m_{\mu}=0.105658$ GeV                    & \cite{Nakamura:2010} \\ \hline  
$\alpha_s(M_Z)=0.1184 \pm 0.0007$         & \cite{Nakamura:2010}     & $\hat \alpha_{em}(M_Z)=1/127.916 $                     & \cite{Nakamura:2010}          \\ 
$\alpha_s(\mu_b)=0.2161$                  &                          &$\hat\alpha_{em}(m_b)=1/133$                               &           \\ 
$\sin^2\hat\theta_W(M_Z)=0.23116(13)$     & \cite{Nakamura:2010}&$G_F/(\hbar c)^3=1.16637(1)\;\textrm{GeV}^{-2}$& \cite{Nakamura:2010}\\ \hline
$ f_B=194 \pm 10$ MeV                          & Table~\ref{tab:lattice}& $\tau_B=1.519 \pm0.007$ ps                             & \cite{Nakamura:2010}          \\
$ f_{B_s} = 234 \pm 10 {\rm MeV}$ & Table~\ref{tab:lattice} & $ \tau_{B_s} = 1.472 \pm 0.026\ {\rm ps}    $ & \cite{Nakamura:2010} \\
 \hline
$f_{K^*,\perp}$(1 GeV)$=0.185 \pm0.009$ GeV  & \cite{Ball:2007}         & $f_{K^*,\parallel}=0.220 \pm0.005$ GeV                 & \cite{Ball:2007}              \\
$a_{1,\perp}$(1 GeV)$=0.10\pm0.07$          & \cite{Ball:2004rg}       & $a_{1,\parallel}$(1 GeV)$=0.10 \pm0.07$                   & \cite{Ball:2004rg}              \\
$a_{2,\perp}$(1 GeV)$=0.13 \pm0.08$          & \cite{Ball:2004rg}       & $a_{2,\parallel}$(1 GeV)$=0.09 \pm0.05$                   & \cite{Ball:2004rg}              \\
$\lambda_{B,+}$(1 GeV)$=0.46 \pm 0.11$ GeV   & \cite{Ball:2006nr}       &  &             \\ \hline
$\mu_b=m_b^{pole}$                        &                          & $\mu_0=2 M_W$                                          &                               \\ 
$\mu_f=\sqrt{0.5 \times \mu_b}$ GeV       & \cite{Beneke:2004dp}     &                                                        &                               \\ \hline
\end{tabular}}
\caption{Input parameters. \label{tab:input}}
\end{center}
\end{table}
%


\section{Results}
\subsection{Separate bounds} 

We first study the individual constraints from the observables described in Section~\ref{sec:obs}.
The main players in our analysis are the radiative decay $\bar B \rightarrow X_s \gamma$, the
leptonic decay $B \rightarrow \mu^+\mu^-$,  and the semileptonic decays $\bar B \rightarrow X_s/K^* \mu^-\mu^-$ 

\begin{figure}[h!]
\begin{center}
\includegraphics[width=7.cm]{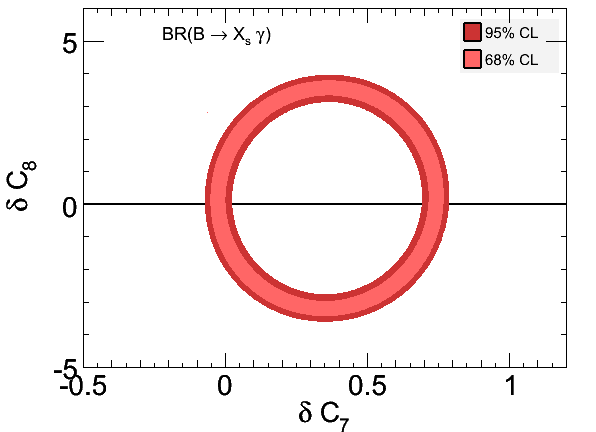}\quad\includegraphics[width=7.cm]{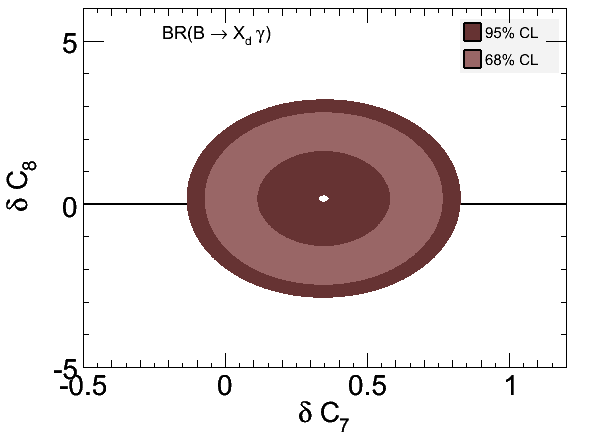}
\caption{68\% and 95\% CL bounds on $\delta C_7$ and $\delta C_8$ induced by the inclusive decays $\bar B\to X_s \gamma$ (left) and  $\bar B \to X_d \gamma$ (right).}
\label{bsgamma}
\end{center}
\end{figure}

Figure~\ref{bsgamma} shows that similar zones are probed by the inclusive decays $\bar B\to X_s \gamma$ and $\bar B \to X_d \gamma$.
The bounds in the ($\delta C_7,\delta C_8$) planes induced by the two inclusive decays are nicely consistent with each other  as expected 
in the MFV framework which predicts a strong correlation between the two decays. Clearly, due to the smaller theoretical and experimental 
error the $\bar B  \to X_s \gamma$  bound is much stronger.

In  the previous MFV analysis~\cite{Hurth:2008jc} the approximation 
was used  that the NP contributions 
to the electromagnetic and chromomagnetic   operators appear in a fixed  linear combination, namely $\delta C_7 + 0.3 \delta C_8$.
This additional assumption  was necessary in the previous analysis due to the limited number of independent experimental measurements. 
The correlations  between $\delta C_7$ and $\delta C_8$, shown  in Figure~\ref{bsgamma},  do not support this simplifying assumption. 

\begin{figure}[h!]
\begin{center}
\includegraphics[width=7.cm]{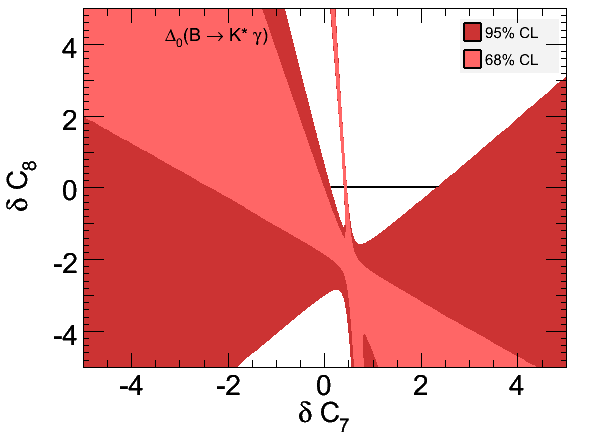}
\caption{68\% and 95\% bounds on   on $\delta C_7$ and $\delta C_8$   induced by the isospin asymmetry in $B\to K^* \gamma$.}
\label{isospin}
\end{center}
\end{figure}

The isospin asymmetry in the exclusive mode $B\to K^* \gamma$ brings complementary information to the inclusive branching ratios.
Figure~\ref{isospin} shows  that the isospin asymmetry seems to favor opposite signs for $\delta C_7$ and $\delta C_8$.

\begin{figure}[h!]
\begin{center}
\includegraphics[width=7.cm]{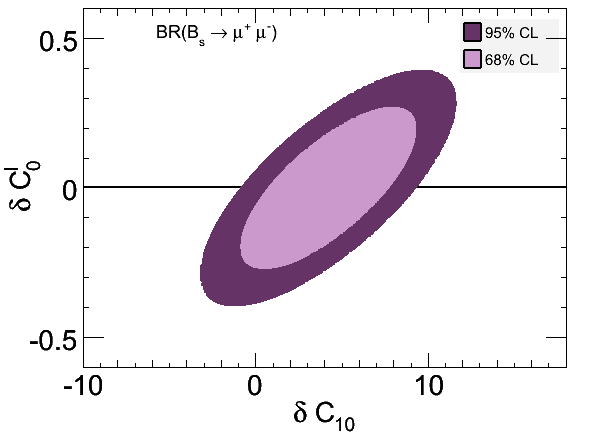}\quad\includegraphics[width=7.cm]{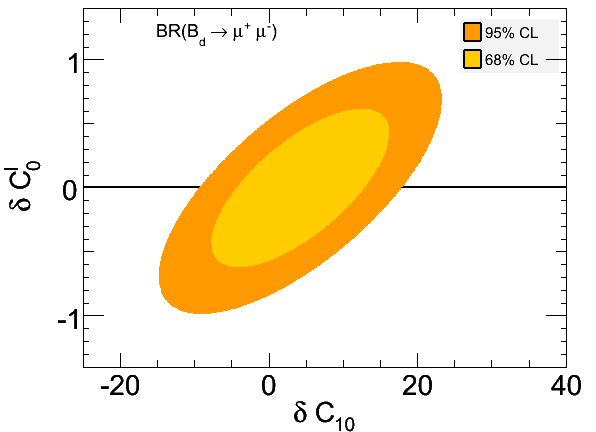}
\caption{68\% and 95\%  CL bounds on  on $\delta C_{10}$ and $\delta C_0^\ell$  induced by the decays $B_s\to \mu^+\mu^-$ (left), $B_d\to \mu^+\mu^-$ (right).}
\label{bll}
\end{center}
\end{figure}

The leptonic decays $B_s\to \mu^+\mu^-$ and $B_d \to \mu^+\mu^-$ are sensitive for $\delta C_{10}$ and the scalar contribution $\delta C_0^\ell$.
The shapes in the corresponding correlation plots  induced by the two leptonic decays are very similar, thus, highly consistent with each other
as can be seen in Figure~\ref{bll}.     
This feature strongly supports  the MFV  hypothesis  which predicts a strong correlation between these two decays as given in Eq.~(\ref{bllcorrelation}).
Of course, the experimental limit for the decay  $B_s  \to \mu^+\mu^-$ is much tighter and therefore the present constraints are much stronger. 
 We therefore take the decay $B_d \to \mu^+\mu^-$  out of the 
global MFV fit and will make  a prediction for this decay within the MFV framework below. 

We notice that the constraint on the scalar coefficient induced by the decay  $B_s\to \mu^+\mu^-$ is very strong and a  large scalar contribution is not 
allowed anymore.

\begin{figure}[h!]
\begin{center}
\includegraphics[width=7.cm]{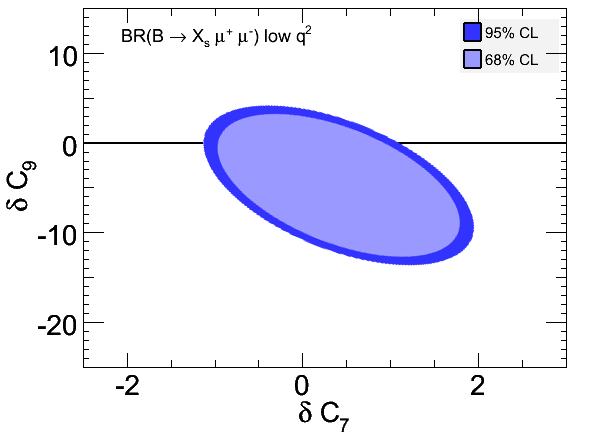}\quad\includegraphics[width=7.cm]{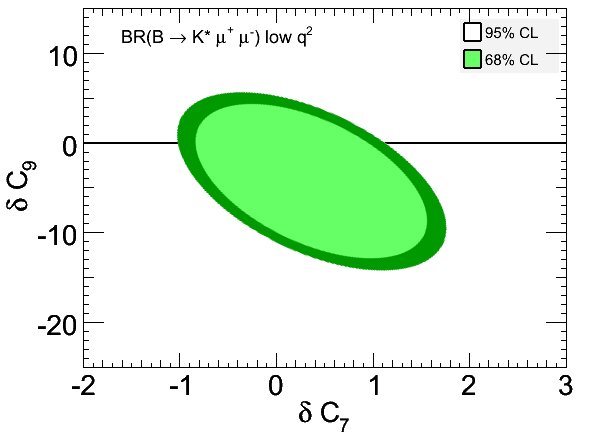}\\[0.3cm]
\includegraphics[width=7.cm]{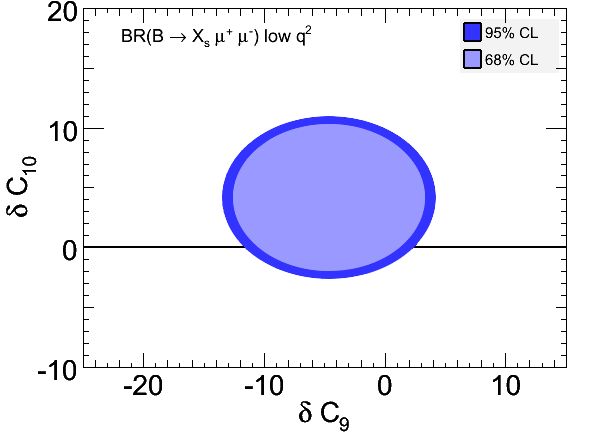}\quad\includegraphics[width=7.cm]{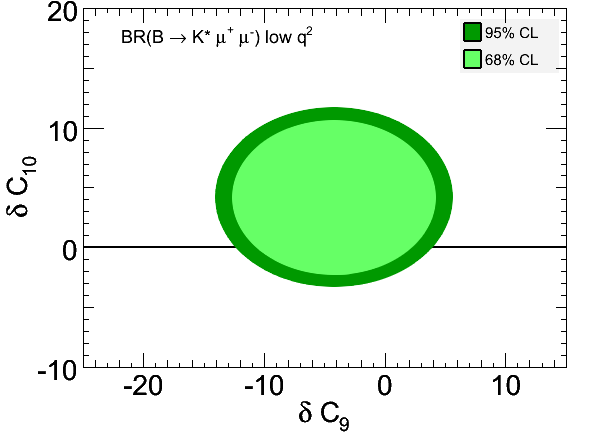}\\[0.3cm]
\includegraphics[width=7.cm]{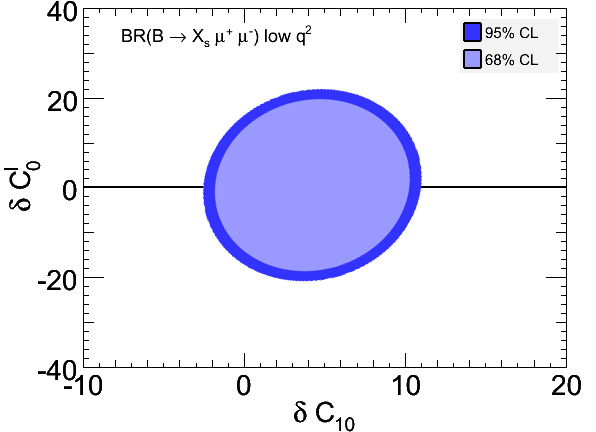}\quad\includegraphics[width=7.cm]{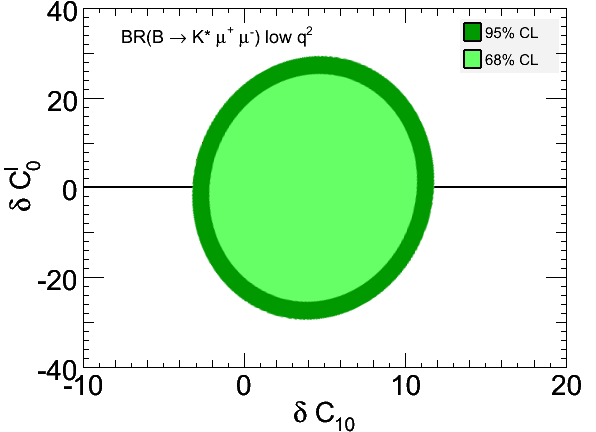}
\caption{68\% and 95\% CL bounds on various $\delta C_i$ induced by the decays $B\to X_s\mu^+\mu^-$ (left) and $B\to K^*\mu^+\mu^-$ (right) at low-$q^2$.}
\label{K8X_s}
\end{center}
\end{figure}

The low-$q^2$ data of the inclusive decay  $\bar B\to X_s\mu^+\mu^-$ and of the exclusive decay $B\to K^*\mu^+\mu^-$ have similar constraining power, as can be 
seen in Figure~\ref{K8X_s}.
It is nontrivial  that the correlation plots of the various $\delta C_i$ look almost identical for the inclusive and the exclusive mode.  
$\bar  B\to X_s\mu^+\mu^-$ has small theoretical errors and large experimental errors, while the situation is reversed for $B\to K^*\mu^+\mu^-$. 
A statistical combination of both allows to enhance their effect. 
However, one realizes the potential of the inclusive mode if one takes into account the fact that the recent Babar and Belle measurements of the inclusive branching 
ratios~\cite{Aubert:2004it, Iwasaki:2005sy}
 only use less than a quarter of the available data sets of the B factories.
 The constraints on $C_{10}$ are similar to those from $B_s\to \mu^+\mu^-$, but contrary to $B_s\to \mu^+\mu^-$, 
the constraints on the scalar contributions here is very weak.

 Finally, we note that the allowed values of $\delta C_9$ and $\delta C_{10}$ are much smaller in specific NP models  than within a model-independent analysis, so for example 
 the structure of the CMSSM already bounds  their values significantly before any experimental data is used (see Ref.~\cite{Mahmoudi:2012un}).

\subsection{Fit results} 
 
\begin{figure}[h!]
\begin{center}
\includegraphics[width=6.5cm]{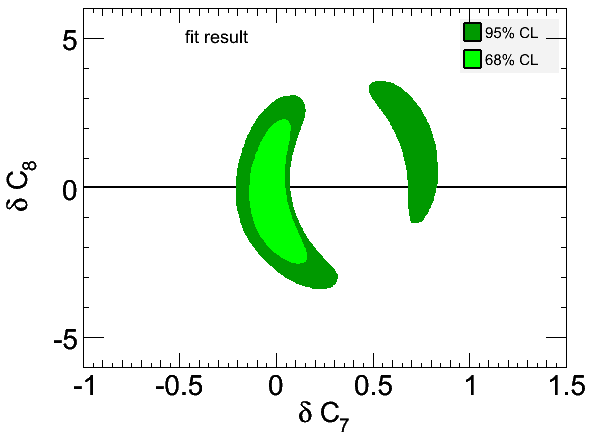}\quad\includegraphics[width=6.5cm]{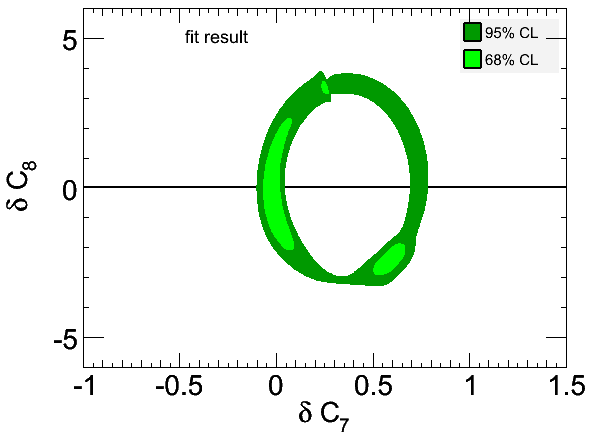}\\[0.3cm]
\includegraphics[width=6.5cm]{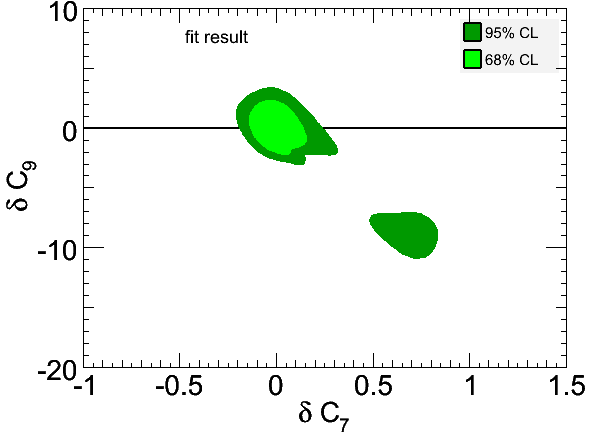}\quad\includegraphics[width=6.5cm]{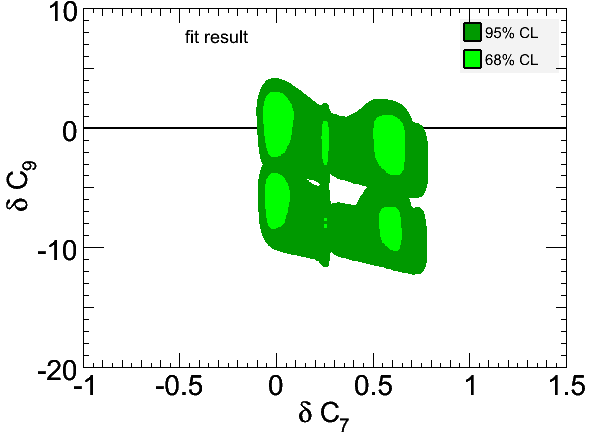}\\[0.3cm]
\includegraphics[width=6.5cm]{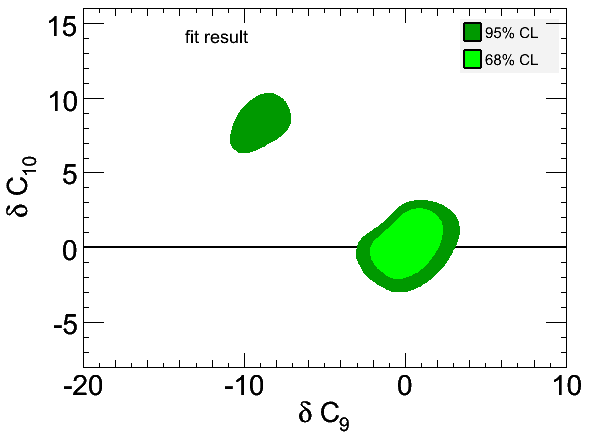}\quad\includegraphics[width=6.5cm]{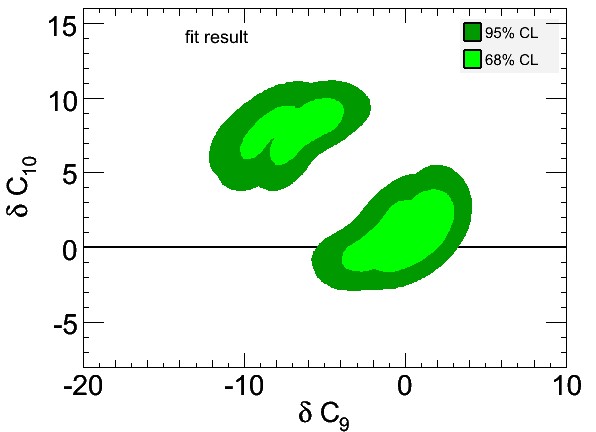}\\[0.3cm]
\includegraphics[width=6.5cm]{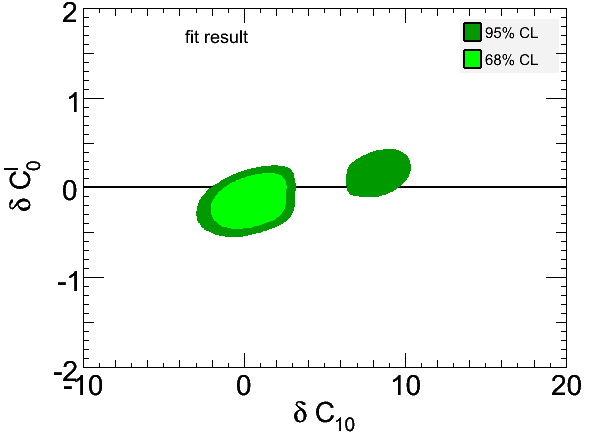}\quad\includegraphics[width=6.5cm]{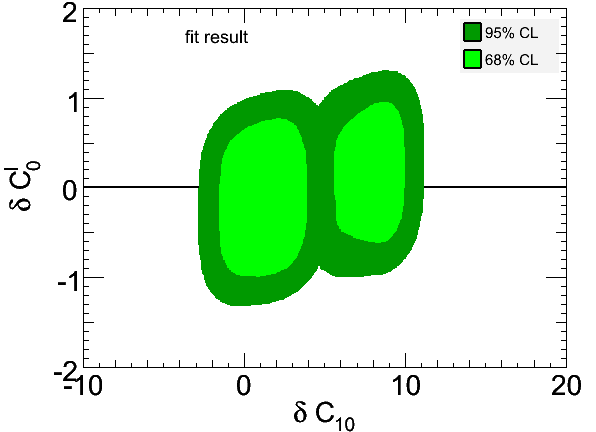}
\caption{Global MFV fit to the various NP coefficients $\delta C_i$ in the MFV effective theory {\it with}  (left) and  {\it without} experimental data of LHCb (right).}
\label{MFVfit}
\end{center}
\end{figure}

We made two global MFV fits in order to make the significance of the latest LHCb data manifest, see Figure~\ref{MFVfit}.
First we have used the experimental data before the start of the LHCb experiment (pre-LHCb, right plots).
These measurements are listed in Table~\ref{tab:obs-preLHCb}.  Then we have included the latest LHCb measurements given in Table~\ref{tab:obs}
(post-LHCb, left plots).  

Here $C_8$ is mostly constrained by $\bar B\to X_{s,d} \gamma$, while $C_7$ is constrained by many other observables as well.
$C_9$ is highly constrained by $b \to s \mu^+\mu^-$ (inclusive and exclusive). $C_{10}$ is in addition further constrained by $B_s\to \mu^+\mu^-$.
$C_0^l$ is dominantly constrained by $B_s\to \mu^+\mu^-$.

There are  always two allowed regions  at 95\% CL in the correlation plots within  the post-LHCb fit; one corresponds to SM-like MFV  coefficients and one to 
coefficients with flipped sign. The allowed region with the SM is more favored. The various $\delta {C_i}$-correlation plots show the flipped-sign for $C_7$
is only possible if $C_9$ and $C_{10}$ receive large non-standard contributions which finally also change the sign of these coefficients.

\begin{figure}[h!]
\begin{center}
\includegraphics[width=6.5cm]{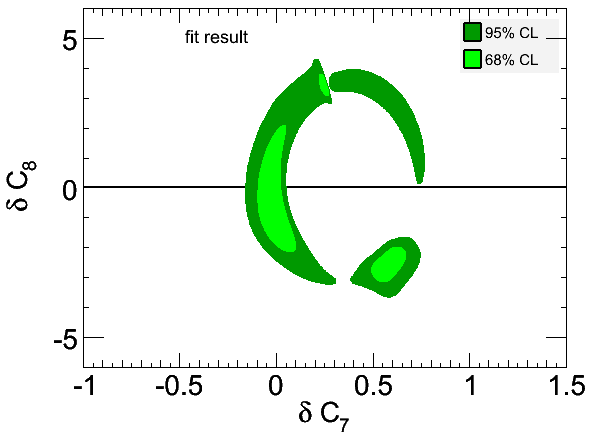}\quad\includegraphics[width=6.5cm]{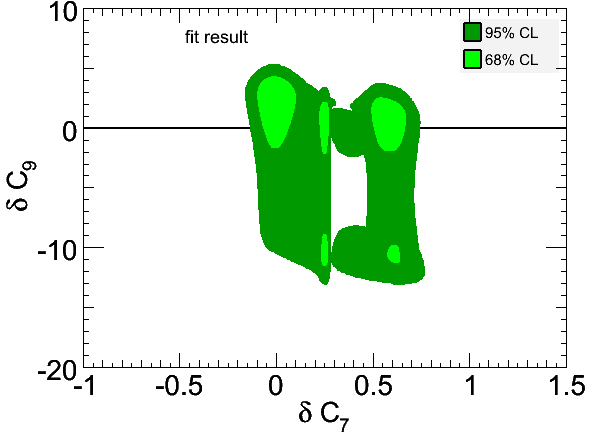}\\[0.3cm]
\includegraphics[width=6.5cm]{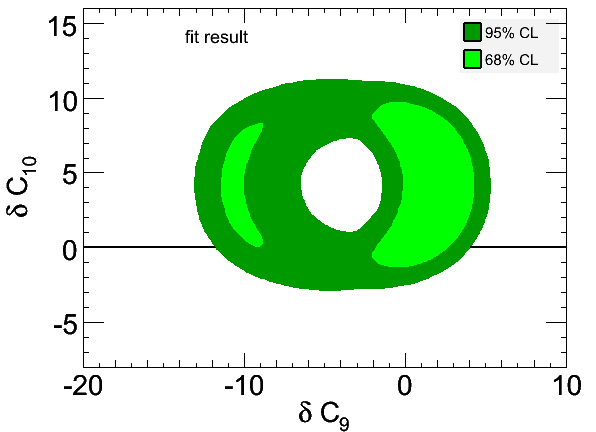}\quad\includegraphics[width=6.5cm]{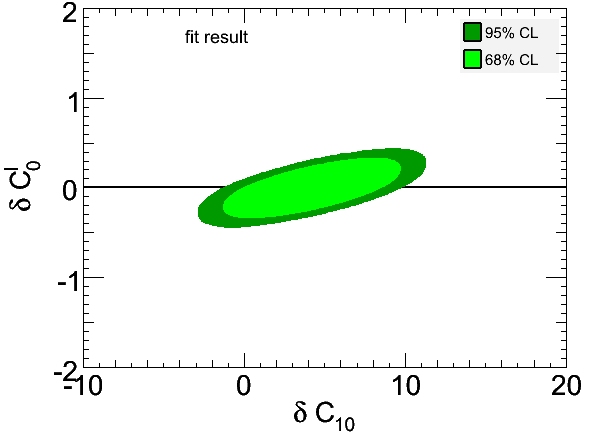}
\caption{Global MFV fit with the latest data set {\it excluding}  all LHCb measurements of $B\to K^* \mu^+\mu^-$ observables.}
\label{MFVfitextra}
\end{center}
\end{figure}

We have also studied the impact of the LHCb measurements of the branching ratio, of the forward-backward asymmetry, and of the $K^*$ polarization within the exclusive decay 
$B \to K^* \ell^+\ell^-$ by taking these LHCb measurements out of the fit.  The results in Table~\ref{MFVfitextra} show that these pieces of experimental information from the LHCb experiment 
are very  important. They significantly reduce the allowed areas for $\delta C_9$ and $ \delta C_{10}$.

\subsection{Predictions within the MFV benchmark}

With the help of the results of the global fit, which restricts the NP contributions $\delta C_i$, we can now derive several interesting predictions 
of observables which are not yet well measured. This analysis allows to spot these observables which still allow for relatively large deviations
from the SM even in the MFV benchmark scenario. 

\begin{itemize}
\item For the branching ratio of the decay $\bar B \to X_s \tau\tau$ we get  the  95\%CL bounds 
\begin{equation}
0.2 \times 10^{-7}  \,< \, {\rm  BR}(\bar B \to X_s \tau^+\tau^-)_{q^2>14.4 \rm{GeV}^2} \,<\,  3.7 \times 10^{-7}\,.
\end{equation}
This has to be compared with the SM prediction 
\begin{equation}{\rm BR}(\bar B \to X_s \tau^+\tau^-)_{q^2>14.4 \rm{GeV}^2} =  (1.61 \pm 0.40)\times 10^{-7}\, .\end{equation}
So there are still  large deviations from  the SM prediction of this observable possible within the MFV scenario.  
And as  stated above, any measurement beyond the MFV  bounds would indicate the existence of new flavour structures.

\item For the zero-crossing of the forward-backward asymmetry  in the inclusive decay $\bar  B \to X_s \mu^+\mu^-$, we get 
the lower bound at the  95\%CL  
\begin{equation}
A_{FB}(q^2_0) = 0\,; \,\,\,\,\,\,   1.94\, {\rm GeV}^2  \, < \, q_0^2\,,    
\end{equation}
while the very precise SM prediction is $(q^2_0)^{\rm SM}  =  (3.40  \pm 0.25)   \rm{GeV}^2$. There is natural upper bound given by the cut 
due to the charm resonances. Due to the theoretical cleanliness of this observable, there are still large deviations from the SM prediction 
possible within the MFV benchmark. This is also true for the the complete function $A_{FB}(q^2)$.

\item We have taken the measurements of the decays $\bar B \to X_d \gamma$ and $B_d \to \mu^+\mu^-$ out of the global fit and find 
the following MFV predictions again at the 95\%CL: 
\begin{equation}
1.0 \times 10^{-5}   \,<\,  {\rm BR}(\bar B \to X_d \gamma) \, < \, 4.0 \times 10^{-5} \,; \, \,\,\,   {\rm BR}(B_d  \to \mu^+\mu^-)\, <\,  3.8  \times 10^{-10}\,.  
\end{equation}
 The corresponding SM predictions are: 
\begin{equation}
{\rm BR}(\bar B \to X_d \gamma)^{\rm SM}  \, =  (1.49 \pm 0.30) \times 10^{-5} \,; \, \,\,\,   {\rm BR}(B_d  \to \mu^+\mu^-)^{\rm SM} \, =\,  (1.11 \pm   0.27)   \times 10^{-10}\,,   
\end{equation}
and the present experimental data (see Table~\ref{Inputobservables}) is:
\begin{equation}
{\rm BR}(\bar B \to X_d \gamma)^{\rm Exp.} \, =  (1.41 \pm 0.57) \times 10^{-5} \,; \, \,\,\,   {\rm BR}(B_d  \to \mu^+\mu^-)^{\rm Exp} \, <\,  10.0  \times 10^{-10}\,.  
\end{equation}
So the present $\bar B \to X_d \gamma$ measurement is already below the MFV bound and is nicely consistent with the correlation between the decays 
$\bar B \to X_s \gamma$ and   $\bar B \to X_d \gamma$ predicted in the MFV scenario. 
In the case of the leptonic decay $B_d \to \mu^+\mu^-$, however, the MFV bound is stronger than the current experimental limit.
And there are still sizable deviations from the SM prediction possible within the MFV  but
an enhancement by orders of magnitudes due to large $\tan \beta$ effects are already ruled out by the latest measurements. 

\item    For the large set of angular $B \to K^* \mu^+\mu^-$ observables discussed in Section~\ref{sec:obs} we also can easily derive their  MFV predictions. In
an exemplary mode,  we give the 95\%CL MFV 
predictions for the $A_T^{(i)}$, averaged  over the low-$q^2$ region ($1\,{\rm  GeV}^2 < q^2 < 6\,{\rm GeV}^2$),
 \begin{equation} 
-0.065  \,<\,    \langle A_T^{(2)}\rangle  \, <\,       -0.022;\,\,\,\,\,\,  0.34  \, <\,   \langle  A_T^{(3)}\rangle   \,<\,   0.99;\,\,\,\,\,\,  0.19     \,<\,   \langle  A_T^{(4)}\rangle \,<\,    1.27\,,
\end{equation} 
and for the $H_T^{(i)}$,  averaged  over the high-$q^2$ region ($14.18\,{\rm  GeV}^2 < q^2 < 16\,{\rm GeV}^2$), 
 \begin{equation} 
\langle H_T^{(1)}\rangle = 1,\,\,\,\,\,\,  -1.01  \, <\,   \langle H_T^{(2)}\rangle   \,<\,   -0.44, \,\,\,\,\,\,\,\, -1.01  \, <\,   \langle H_T^{(3)}\rangle   \,<\,   -0.44.
\end{equation} 
Due to the experimental and theoretical uncertainties of  the $A_T^{(i)}$ observables, discussed in Refs.~\cite{Egede:2008uy} or \cite{Beaujean:2012uj},
the predicted MFV range cannot be really separated from the SM prediction; any significant deviation from the SM prediction indicates new flavour structures. But
for the $H_T^{(i)}$ observables   deviations form the SM are still
possible within the MFV scenario.

\item For the rare $s \to d$ transitions we refer again to the previous analysis in Ref.~\cite{Hurth:2008jc}. 
We just state that   the existing  rather weak experimental bound on the branching ratio of ${\rm BR}(K^+ \to \pi^+ \nu\bar\nu)$~\cite{Adler:2008zz}
implies MFV predictions for the flavour observables ${\rm BR}(B\to K^{(*)} \nu \bar \nu)$ and ${\rm BR}(K_L\to\pi^0\nu\bar\nu)$. Moreover, 
as in the case of the two decays $B_{s,d}\to \mu\mu$, the charged and neutral $K \to \pi \nu\bar\nu$ decays are governed by only one parameter,
namely the real coefficient $\delta C_{\nu\bar\nu}$  in the MFV effective  theory  (see Eq.(\ref{NUCoeff})), thus, the ratio of the two $K \to \pi \nu\bar\nu$ allows for an important model-independent test 
of the MFV hypothesis. There will be  two dedicated kaon experiments,  E-14 Koto at J-PARC~\cite{E14}  and Na42 at CERN~\cite{Na62}, for this task in the near future.
\end{itemize}

\section*{Acknowledgement} TH thanks  the CERN theory group for its  hospitality during his regular visits to CERN where
part of this work  was written \\


\begin{appendix}
\numberwithin{equation}{section}
\section{Effective, new, and modified  Wilson coefficients}\label{A:appendix}

{\it  The standard effective Wilson coefficients}  are defined as follows:
\begin{equation} \label{effective}
C_i^{\rm eff}(\mu)  =   C_i(\mu),  \mbox{ for $i = 1, ..., 6$,}\,\,\,\, C_7^{\rm eff}(\mu)  =  C_7(\mu) + \sum_{j=1}^6 y_j C_j(\mu),\,\,\,\, C_8^{\rm eff}(\mu)   =  C_8(\mu) + \sum_{j=1}^6 z_j C_j(\mu).
\end{equation}
In the $\overline{\rm MS}$ scheme, one fixes  $\vec{y} = (0, 0,
-\frac{1}{3}, -\frac{4}{9}, -\frac{20}{3}, -\frac{80}{9})$ and $\vec{z} = (0, 0, 1, -\frac{1}{6},
20, -\frac{10}{3})$, then the leading-order $b \to s
\gamma$ and $b \to sg$ matrix elements of the effective Hamiltonian are
proportional to the leading-order terms in $C_7^{\rm eff}$ and
$C_8^{\rm eff}$~\cite{Misiak:2006ab}. Moreover, we have  $C_{10}^{\rm eff} = C_{10}(\mu)$, and $C_9^{\rm eff}$ is defined as:
\begin{eqnarray}
C_9^{\rm eff}(s) &=&  C_9(\mu)
+ \sum_{i=1}^6 C_i(\mu) \gamma^{(0)}_{i9}  \ln\left(\frac{m_b}{\mu}\right)
+ \frac{4}{3} C_3(\mu)+ \frac{64}{9} C_5(\mu)+ \frac{64}{27} C_6(\mu)
\nonumber\\
&+& g\left(\hat{m}_c,s \right) \left( 
\frac{4}{3} C_1(\mu) + C_2(\mu)   
+ 6 C_3(\mu) + 60 C_5(\mu) \right)
\nonumber\\ &+& g(1,s) \left(  
-\frac{7}{2} C_3(\mu)-\frac{2}{3} C_4(\mu)-38 C_5(\mu)-\frac{32}{3} C_6(\mu) \right)
\nonumber\\ &+& g(0,s) \left(  
-\frac{1}{2} C_3(\mu)-\frac{2}{3} C_4(\mu)- 8 C_5(\mu)-\frac{32}{3} C_6(\mu) \right)
\label{effWilson}
\end{eqnarray}
where the function $g(\hat{m}_c,s)$  is given as
 \begin{eqnarray}
g(z,s) &=&  - \frac{4}{9} \ln(z) + \frac{8}{27} + \frac{16}{9}\frac{z}{s} 
             - \frac{2}{9} \left( 2+\frac{4\, z}{s} \right)
             \sqrt{\left|\frac{4\,z-s}{s}\right|} \times \nonumber\\
       &\times& \{\begin{array}{ll}
 2 \arctan \sqrt{\frac{s}{4\,z-s}} & \mbox{for} \, s < 4\,z\, ,   \\ 
 \ln \left(\frac{\sqrt{s} + \sqrt{s - 4\,z}}{\sqrt{s} - 
\sqrt{s - 4\,z}} \right) -i\,\pi \qquad &\mbox{for}\,  s > 4\,z  \, . \end{array} \right.
\end{eqnarray}

{\it  The {new} Wilson coefficients}  are introduced  in Ref.~\cite{Ghinculov:2003qd} and are defined as:  
\begin{eqnarray}
  C^{new}_7(s) &=&  \left(1+\frac{\alpha_s}{\pi}\sigma_7 (s)\right) 
       C_7^{\rm eff} 
        -\frac{\alpha_s}{4\,\pi} \left[ C_1^{(0)} F_1^{(7)}(s)+
        C_2^{(0)} F_2^{(7)}(s) 
       + C^{\rm eff(0)}_8 F_8^{(7)}(s) \right]\;, \\
  C^{new}_9(s)  &=&  \left(1+\frac{\alpha_s}{\pi} \sigma_9 (s) \right) 
        C_9^{\rm eff}(s) 
       -\frac{\alpha_s}{4\,\pi} \left[ C_1^{(0)} F_1^{(9)}(s) 
        + C_2^{(0)} F_2^{(9)}(s)+ C_8^{\rm eff (0)} F_8^{(9)}(s) \right]\;, \\
  C^{new}_{10}(s)  &=& \left( 1+\frac{\alpha_s}{\pi} 
        \sigma_{9} (s) \right) C^{\rm eff}_{10}\, . 
\end{eqnarray}
The virtual corrections to $O_{1,2}$ and $O_8$ are embedded in $F_{1,2}^{(7,9)}$ and $F_8^{(7,9)}$. The $\sigma_i$ functions  indicate certain bremsstrahlung contributions
(see Ref.~\cite{Ghinculov:2003qd}).

Finally, {\it the modified effective Wilson coefficients}  introduced in Ref.~\cite{Grinstein:2004vb}  include 
contributions of the four-quark operators but {\it also} of the gluon  dipole operators. It is important to note that
these quantities are different from {\it  the effective Wilson coefficients}  $\wilson[eff]{7,9}$ introduced before. 
\begin{align}  
  \label{eq:c9effGP}
  \wilson[eff,mod]{9} & = 
    \wilson{9} + 
    h(0, q^2) \[ \frac{4}{3}\, \wilson{1} + \wilson{2} + \frac{11}{2}\, \wilson{3}
      - \frac{2}{3}\, \wilson{4} + 52\, \wilson{5} - \frac{32}{3}\, \wilson{6}\] 
\\
  & - \frac{1}{2}\, h(m_b, q^2) \[ 7\, \wilson{3} + \frac{4}{3}\, \wilson{4} + 76\, \wilson{5}
      + \frac{64}{3}\, \wilson{6} \]
    + \frac{4}{3} \[ \wilson{3} + \frac{16}{3}\, \wilson{5} + \frac{16}{9}\, \wilson{6} \]
\nonumber\\
  & + \frac{\alpha_s}{4 \pi} \[ \wilson{1} \left(B(q^2) + 4\, C(q^2)\right) 
      - 3\, \wilson{2}\left(2\, B(q^2) - C(q^2)\right) - \wilson{8}F_8^{(9)}(q^2) \]  
\nonumber\\ 
  & + 8 {\frac{m_c^2}{q^2}  \[\frac{4}{9}\,\wilson{1} 
      + \frac{1}{3}\,\wilson{2} + 2\,\wilson{3} + 20\,\wilson{5} \] },  
\nonumber \\
  \label{eq:c7effGP}
  \wilson[eff,mod]{7} & = 
  \wilson{7} - 
  \frac{1}{3} \[ \wilson{3} + \frac{4}{3}\,\wilson{4} + 20\,\wilson{5} 
    + \frac{80}{3}\wilson{6} \]
  + \frac{\alpha_s}{4 \pi} \[ \left(\wilson{1} - 6\,\wilson{2}\right) A(q^2) 
    - \wilson{8} F_8^{(7)}(q^2)\] ,
\end{align}
The functions $A,B,C$ and $F_8^{(7)}, F_8^{(9)}$  are given  in  Refs.~\cite{Seidel:2004jh} and \cite{Beneke:2001at}.

\section{Determination of the soft form factors}

To obtain the soft  $B \to K^*$ form factors  we have used the following  factorization scheme~ \cite{Beneke:2004dp}:
\begin{align}
  \xi_\perp (q^2) & = \frac{M_B}{M_B + m_{K^*}} V(q^2)\;,\\
  \xi_\parallel (q^2) & = \frac{M_B + m_{K^*}}{2 E_{K^*}} A_1 (q^2) -\frac{M_B - m_{K^*}}{M_B} A_2 (q^2)\;.
\end{align}
The full form factors $V$ and $A_{1,2}$ have been taken from light-cone sum rule (LCSR) calculations \cite{Ball:2004rg}:
\begin{align}
V(q^2) &= \frac{r_1}{1 - q^2/m_R^2} + \frac{r_2}{1 - q^2/m_{fit}^2} \;,\\ 
A_1(q^2) &= \frac{r_2}{1 - q^2/m_{fit}^2} \;,\\ 
A_2(q^2) &= \frac{r_1}{1 - q^2/m_{fit}^2} + \frac{r_2}{(1 - q^2/m_{fit}^2)^2} \;,
\label{eq:qdependent-ff}
\end{align}
where the fit parameters $r_{1,2}, m^2_{R}$ and $m^2_{fit}$ are given in Table \ref{tab:FF:fit}.
\begin{table}[t]
\centering
\begin{tabular}{|c|cccc|}
\hline
  $ $ & $r_1$ & $r_2$ & $m_R^2\,[\text{GeV}^2]$ & $m_{fit}^2\,[\text{GeV}^2]$ \\
\hline
  $V$   & $0.923$  & $-0.511$ & $5.32^2$ & $49.40$ \\[0.5ex]
  $A_1$ &  \mbox{} & $ 0.290$ &  \mbox{} & $40.38$ \\[0.5ex]
  $A_2$ & $-0.084$ &  $0.342$ &  \mbox{} & $52.00$ \\[0.5ex]
\hline
\end{tabular}
\caption{\label{tab:FF:fit} Fit parameters describing the $q^2$ dependence
 of the form factors $V$ and $A_{1,2}$ in the LCSR approach \cite{Ball:2004rg}.}
\end{table}

\end{appendix}



\begin{thebibliography}{12}



\bibitem{Belle} 
  Belle collaboration: http://belle.kek.jp/

\bibitem{Babar} 
  BaBar collaboration: http://www.slac.stanford.edu/BFROOT/


\bibitem{TevatronB1}
  CDF collaboration: http://www-cdf.fnal.gov/physics/new/bottom/bottom.html

\bibitem{TevatronB2}
  D0 collaboration: http://www-d0.fnal.gov/Run2Physics/WWW/results/b.htm



\bibitem{LHCb}  LHCb collaboration:  http://lhcb.web.cern.ch/lhcb/



\bibitem{Kobayashi:1973fv}
  M.~Kobayashi and T.~Maskawa,
  Prog.\ Theor.\ Phys.\  {\bf 49} (1973) 652.


\bibitem{Cabibbo:1963yz}
  N.~Cabibbo,
  Phys.\ Rev.\ Lett.\  {\bf 10} (1963) 531.



\bibitem{Hurth:2010tk}
  T.~Hurth and M.~Nakao,
  Ann.\ Rev.\ Nucl.\ Part.\ Sci.\  {\bf 60} (2010) 645
  [arXiv:1005.1224 [hep-ph]].



\bibitem{Hurth:2003vb}
  T.~Hurth,
  Rev.\ Mod.\ Phys.\  {\bf 75} (2003) 1159
  [hep-ph/0212304].



\bibitem{Mahmoudi:2012uk}
  F.~Mahmoudi,
  arXiv:1205.3099 [hep-ph].

\bibitem{Hurth:2011zy}
  T.~Hurth and S.~Kraml,
  AIP Conf.\ Proc.\  {\bf 1441} (2012) 713
  [arXiv:1110.3804 [hep-ph]].


\bibitem{Chivukula:1987py}
  R.~S.~Chivukula and H.~Georgi,
  Phys.\ Lett.\ B {\bf 188} (1987) 99.



\bibitem{Hall:1990ac}
  L.~J.~Hall and L.~Randall,
  Phys.\ Rev.\ Lett.\  {\bf 65}, 2939 (1990).



\bibitem{D'Ambrosio:2002ex} 
G.~D'Ambrosio, G.~F.~Giudice, G.~Isidori and A.~Strumia,
Nucl.\ Phys.\ B {\bf 645} (2002)  155
[hep-ph/0207036].





\bibitem{Hurth:2008jc}
  T.~Hurth, G.~Isidori, J.~F.~Kamenik and F.~Mescia,
  Nucl.\ Phys.\ B {\bf 808} (2009) 326
  [arXiv:0807.5039 [hep-ph]].



\bibitem{Beneke:2001at}
  M.~Beneke, T.~Feldmann and D.~Seidel,
  Nucl.\ Phys.\ B {\bf 612} (2001) 25
  [hep-ph/0106067].




\bibitem{Beneke:2004dp}
  M.~Beneke, T.~.Feldmann and D.~Seidel,
  Eur.\ Phys.\ J.\ C {\bf 41} (2005) 173
  [hep-ph/0412400].




\bibitem{Grinstein:2004vb}
  B.~Grinstein and D.~Pirjol,
  Phys.\ Rev.\ D {\bf 70} (2004) 114005
  [hep-ph/0404250].




\bibitem{Beylich:2011aq}
  M.~Beylich, G.~Buchalla and T.~Feldmann,
  Eur.\ Phys.\ J.\ C {\bf 71} (2011) 1635
  [arXiv:1101.5118 [hep-ph]].



\bibitem{Bona:2005eu}
  M.~Bona {\it et al.}  [UTfit Collaboration],
  JHEP {\bf 0603} (2006) 080
  [hep-ph/0509219].



\bibitem{Lenz:2010gu}
  A.~Lenz, U.~Nierste, J.~Charles, S.~Descotes-Genon, A.~Jantsch, C.~Kaufhold, \
H.~Lacker and S.~Monteil {\it et al.},
  Phys.\ Rev.\ D {\bf 83} (2011) 036004
  [arXiv:1008.1593 [hep-ph]].


\bibitem{Lenz:2012az}
  A.~Lenz, U.~Nierste, J.~Charles, S.~Descotes-Genon, H.~Lacker, S.~Monteil, V.\
~Niess and S.~T'Jampens,
  arXiv:1203.0238 [hep-ph].













\bibitem{Mercolli:2009ns}
  L.~Mercolli and C.~Smith,
  Nucl.\ Phys.\ B {\bf 817} (2009) 1
  [arXiv:0902.1949 [hep-ph]].



\bibitem{Paradisi:2009ey}
  P.~Paradisi and D.~M.~Straub,
  Phys.\ Lett.\ B {\bf 684} (2010) 147
  [arXiv:0906.4551 [hep-ph]].

\bibitem{Feldmann:2006jk}
  T.~Feldmann and T.~Mannel,
  hep-ph/0611095.




\bibitem{Agashe:2005hk}
  K.~Agashe, M.~Papucci, G.~Perez and D.~Pirjol,
  hep-ph/0509117.



\bibitem{Grinstein:2006cg}
  B.~Grinstein, V.~Cirigliano, G.~Isidori and M.~B.~Wise,
  hep-ph/0608123.


\bibitem{Nikolidakis:2007fc}
  E.~Nikolidakis and C.~Smith,
  Phys.\ Rev.\ D {\bf 77} (2008) 015021
  [arXiv:0710.3129 [hep-ph]].


\bibitem{Csaki:2011ge} 
  C.~Csaki, Y.~Grossman and B.~Heidenreich,
  Phys.\ Rev.\ D {\bf 85}, 095009 (2012)
  [arXiv:1111.1239 [hep-ph]].




\bibitem{Paradisi:2008qh}
  P.~Paradisi, M.~Ratz, R.~Schieren and C.~Simonetto,
  Phys.\ Lett.\ B {\bf 668} (2008) 202
  [arXiv:0805.3989 [hep-ph]].


\bibitem{Colangelo:2008qp}
  G.~Colangelo, E.~Nikolidakis and C.~Smith,
  Eur.\ Phys.\ J.\ C {\bf 59} (2009) 75
  [arXiv:0807.0801 [hep-ph]].




\bibitem{Isidori:2012ts}
  G.~Isidori and D.~M.~Straub,
  arXiv:1202.0464 [hep-ph].

\bibitem{Nikolidakis:2007fc}
  E.~Nikolidakis and C.~Smith,
  Phys.\ Rev.\ D {\bf 77} (2008) 015021
  [arXiv:0710.3129 [hep-ph]].


\bibitem{Csaki:2011ge} 
  C.~Csaki, Y.~Grossman and B.~Heidenreich,
  Phys.\ Rev.\ D {\bf 85}, 095009 (2012)
  [arXiv:1111.1239 [hep-ph]].






\bibitem{Antonelli:2008jg}
M.~Antonelli {\it et al.} [FlaviaNet Working Group on Kaon Decays],
arXiv:0801.1817 [hep-ph] and online update at
\verb"http://www.lnf.infn.it/wg/vus/" from Rare K decays and Decay Constants.


\bibitem{Brod:2008ss}
J.~Brod and M.~Gorbahn,
arXiv:0805.4119 [hep-ph].

\bibitem{Buras:2006gb}
A.~J.~Buras, M.~Gorbahn, U.~Haisch and U.~Nierste,
JHEP {\bf 0611}, 002 (2006)
[arXiv:hep-ph/0603079];

\bibitem{Mescia:2007kn}
F.~Mescia and C.~Smith,
Phys.\ Rev.\ D {\bf 76}, 034017 (2007)
[arXiv:0705.2025 [hep-ph]].


  
  

\bibitem{Czarnecki:1998tn}
  A.~Czarnecki and W.~J.~Marciano,
  Phys.\ Rev.\ Lett.\  {\bf 81} (1998) 277
  [hep-ph/9804252].



\bibitem{Misiak:2006zs}
  M.~Misiak, H.~M.~Asatrian, K.~Bieri, M.~Czakon, A.~Czarnecki, T.~Ewerth, A.~Ferroglia and P.~Gambino {\it et al.},
  Phys.\ Rev.\ Lett.\  {\bf 98} (2007) 022002
  [hep-ph/0609232].


\bibitem{Gambino:2008fj}
  P.~Gambino and P.~Giordano,
  Phys.\ Lett.\ B {\bf 669} (2008) 69
  [arXiv:0805.0271 [hep-ph]].


\bibitem{Benzke:2010js}
  M.~Benzke, S.~J.~Lee, M.~Neubert and G.~Paz,
  JHEP {\bf 1008} (2010) 099
  [arXiv:1003.5012 [hep-ph]].



\bibitem{Misiak:2008ss}
  M.~Misiak,
  arXiv:0808.3134 [hep-ph].



\bibitem{Misiak:2006ab}
  M.~Misiak and M.~Steinhauser,
  Nucl.\ Phys.\  B {\bf 764}, 62 (2007)
  [arXiv:hep-ph/0609241].


\bibitem{Hurth:2003dk}
  T.~Hurth, E.~Lunghi and W.~Porod,
  Nucl.\ Phys.\  B {\bf 704} (2005) 56
  [arXiv:hep-ph/0312260].



\bibitem{Kagan:2001zk}
  A.~L.~Kagan and M.~Neubert,
  Phys.\ Lett.\  B {\bf 539}, 227 (2002)
  [arXiv:hep-ph/0110078].



\bibitem{Buchalla:1995vs}
  G.~Buchalla, A.~J.~Buras and M.~E.~Lautenbacher,
  Rev.\ Mod.\ Phys.\  {\bf 68} (1996) 1125
  [hep-ph/9512380].




\bibitem{Bosch:2001gv}
  S.~W.~Bosch and G.~Buchalla,
  Nucl.\ Phys.\ B {\bf 621} (2002) 459
  [hep-ph/0106081].




\bibitem{Ghinculov:2003qd}
  A.~Ghinculov, T.~Hurth, G.~Isidori and Y.~P.~Yao,
  Nucl.\ Phys.\ B {\bf 685} (2004) 351
  [hep-ph/0312128].



\bibitem{Huber:2007vv}
  T.~Huber, T.~Hurth and E.~Lunghi,
  Nucl.\ Phys.\ B {\bf 802} (2008) 40
  [arXiv:0712.3009 [hep-ph]].



\bibitem{Grossman:1996qj}
  Y.~Grossman, Z.~Ligeti and E.~Nardi,
  Phys.\ Rev.\ D {\bf 55} (1997) 2768
  [hep-ph/9607473].


\bibitem{Xiong:2000cp}
  Z.~-h.~Xiong and J.~M.~Yang,
  Nucl.\ Phys.\ B {\bf 602} (2001) 289
  [hep-ph/0012217].



\bibitem{Xiong:2001up}
  Z.~Xiong and J.~M.~Yang,
  Nucl.\ Phys.\ B {\bf 628} (2002) 193
  [hep-ph/0105260].


\bibitem{Ligeti:2007sn}
 Z.~Ligeti and F.~J.~Tackmann,
     Phys.\ Lett.\  B {\bf 653} (2007) 404
   [arXiv:0707.1694v2].



\bibitem{Egede:2008uy}
  U.~Egede, T.~Hurth, J.~Matias, M.~Ramon and W.~Reece,
  JHEP {\bf 0811} (2008) 032
  [arXiv:0807.2589 [hep-ph]].


\bibitem{Kruger:2005ep}
  F.~Kruger and J.~Matias,
  Phys.\ Rev.\ D {\bf 71} (2005) 094009
  [hep-ph/0502060].
.


\bibitem{Kruger:1999xa}
  F.~Kruger, L.~M.~Sehgal, N.~Sinha and R.~Sinha,
  Phys.\ Rev.\ D {\bf 61} (2000) 114028
   [Erratum-ibid.\ D {\bf 63} (2001) 019901]
  [hep-ph/9907386].



\bibitem{Altmannshofer:2008dz}
  W.~Altmannshofer, P.~Ball, A.~Bharucha, A.~J.~Buras, D.~M.~Straub and M.~Wick,
  JHEP {\bf 0901} (2009) 019
  [arXiv:0811.1214 [hep-ph]].




\bibitem{Egede:2010zc}
  U.~Egede, T.~Hurth, J.~Matias, M.~Ramon and W.~Reece,
  JHEP {\bf 1010} (2010) 056
  [arXiv:1005.0571 [hep-ph]].




\bibitem{Bobeth:2008ij}
  C.~Bobeth, G.~Hiller and G.~Piranishvili,
  JHEP {\bf 0807} (2008) 106
  [arXiv:0805.2525 [hep-ph]].


\bibitem{Bobeth:2010wg}
  C.~Bobeth, G.~Hiller and D.~van Dyk,
  JHEP {\bf 1007} (2010) 098
  [arXiv:1006.5013 [hep-ph]].


\bibitem{Bobeth:2011gi}
  C.~Bobeth, G.~Hiller and D.~van Dyk,
  JHEP {\bf 1107} (2011) 067
  [arXiv:1105.0376 [hep-ph]].


\bibitem{Beaujean:2012uj}
  F.~Beaujean, C.~Bobeth, D.~van Dyk and C.~Wacker,
  arXiv:1205.1838 [hep-ph].




\bibitem{Charles:1998dr}
  J.~Charles, A.~Le Yaouanc, L.~Oliver, O.~Pene and J.~C.~Raynal,
  Phys.\ Rev.\ D {\bf 60} (1999) 014001
  [hep-ph/9812358].







\bibitem{Dimopoulos:2011gx}
  P.~Dimopoulos {\it et al.}  [ETM Collaboration],
  {\it Lattice QCD determination of $m_b$, $f_B$ and $f_{B_s}$ with twisted mass Wilson fermions},
  JHEP {\bf 1201} (2012) 046
  [arXiv:1107.1441].


\bibitem{Bazavov:2011aa}
  A.~Bazavov {\it et al.}  [Fermilab Lattice and MILC Collaboration],
  {\it $B$- and $D$-meson decay constants from three-flavor lattice QCD},
  arXiv:1112.3051 [hep-lat].


\bibitem{Neil:2011ku}
  E.~T.~Neil {\it et al.}  [for the Fermilab Lattice and for the MILC Collaborations],
  {\it $B$ and $D$ meson decay constants from 2+1 flavor improved staggered simulations},
  arXiv:1112.3978 [hep-lat].


\bibitem{Na:2012kp}
  H.~Na, C.~J.~Monahan, C.~T.~H.~Davies, R.~Horgan, G.~P.~Lepage and J.~Shigemitsu,
  {\it The $B$ and $B_s$ Meson Decay Constants from Lattice QCD},
  arXiv:1202.4914 [hep-lat].


\bibitem{McNeile:2011ng}
  C.~McNeile, C.~T.~H.~Davies, E.~Follana, K.~Hornbostel and G.~P.~Lepage,
  {\it High-Precision $f_{B_s}$ and HQET from Relativistic Lattice QCD},
  Phys.\ Rev.\ D {\bf 85} (2012) 031503
  [arXiv:1110.4510].


\bibitem{Davies:2012qf}
  C.~Davies,
  {\it Standard Model Heavy Flavor physics on the Lattice},
  arXiv:1203.3862 [hep-lat].

\bibitem{Nakamura:2010}
  K.~Nakamura {\it et al.}  [Particle Data Group],
  {\it Review of particle physics},
  J.\ Phys.\ G {\bf 37} (2010) 075021 and 2011 partial update for the 2012 edition.


\bibitem{Ball:2007}
  P.~Ball, V.~M.~Braun and A.~Lenz,
  {\it Twist-4 distribution amplitudes of the $K^*$ and $\phi$ mesons in QCD},
  JHEP {\bf 0708} (2007) 090
  [arXiv:0707.1201].


\bibitem{Ball:2004rg}
  P.~Ball and R.~Zwicky,
  {\it $B_{d,s} \to \rho, \omega, K^*, \phi$ decay form-factors from light-cone sum rules revisited},
  Phys.\ Rev.\ D {\bf 71} (2005) 014029
  [hep-ph/0412079].


\bibitem{Ball:2006nr}
  P.~Ball and R.~Zwicky,
  {\it $|V_{td} / V_{ts}|$ from $B \to V \gamma$},
  JHEP {\bf 0604} (2006) 046
  [hep-ph/0603232].



\bibitem{Mahmoudi:2007vz}
  F.~Mahmoudi,
  {\it SuperIso: A Program for calculating the isospin asymmetry of $B \to K^* \gamma$ in the MSSM},
  Comput.\ Phys.\ Commun.\  {\bf 178} (2008) 745
  [arXiv:0710.2067].


\bibitem{Mahmoudi:2008tp}
  F.~Mahmoudi,
  {\it SuperIso v2.3: A Program for calculating flavor physics observables in Supersymmetry},
  Comput.\ Phys.\ Commun.\  {\bf 180} (2009) 1579
  [arXiv:0808.3144].


\bibitem{Asner:2010qj}
  D.~Asner {\it et al.}  [Heavy Flavor Averaging Group Collaboration],
  arXiv:1010.1589 [hep-ex]
 and online updates at {\tt http://www.slac.stanford.edu/xorg/hfag}.


\bibitem{delAmoSanchez:2010ae}
  P.~del Amo Sanchez {\it et al.}  [BABAR Collaboration],
  Phys.\ Rev.\ D {\bf 82} (2010) 051101
  [arXiv:1005.4087 [hep-ex]].


\bibitem{Wang:2011sn} 
  W.~Wang,
  arXiv:1102.1925 [hep-ex].


\bibitem{Aaij:2012ac}
  R.~Aaij {\it et al.}  [LHCb Collaboration],
  {\it Strong constraints on the rare decays $B_s \to \mu^+ \mu^-$ and $B^0 \to \mu^+ \mu^-$},
  arXiv:1203.4493 [hep-ex].

\bibitem{LHCb-CONF-2012-008}
 [LHCb Collaboration], 
{\it Differential branching fraction and angular analysis of the $B^0 \to K^{*} \mu^+\mu^-$ decay}
LHCb-CONF-2012-008, presented at the 47th Rencontres de Moriond on
QCD and High Energy Interactions.


\bibitem{Aubert:2004it}
  B.~Aubert {\it et al.}  [BABAR Collaboration],
  Phys.\ Rev.\ Lett.\  {\bf 93} (2004) 081802
  [hep-ex/0404006].


\bibitem{Iwasaki:2005sy}
  M.~Iwasaki {\it et al.}  [Belle Collaboration],
  Phys.\ Rev.\ D {\bf 72} (2005) 092005
  [hep-ex/0503044].


\bibitem{Aaltonen:2007ad}
  T.~Aaltonen {\it et al.}  [CDF Collaboration],
  Phys.\ Rev.\ Lett.\  {\bf 100} (2008) 101802
  [arXiv:0712.1708 [hep-ex]].

\bibitem{CDF}
CDF Collaboration, CDF note 10047.




\bibitem{Mahmoudi:2012un} 
  F.~Mahmoudi, S.~Neshatpour and J.~Orloff,
  arXiv:1205.1845 [hep-ph].


\bibitem{deBruyn:2012wj} 
  K.~de Bruyn, R.~Fleischer, R.~Knegjens, P.~Koppenburg, M.~Merk and N.~Tuning,
  arXiv:1204.1735 [hep-ph].


\bibitem{deBruyn:2012wk} 
  K.~de Bruyn, R.~Fleischer, R.~Knegjens, P.~Koppenburg, M.~Merk, A.~Pellegrino and N.~Tuning,
  arXiv:1204.1737 [hep-ph].


\bibitem{Adler:2008zz}
  S.~Adler {\it et al.}  [E787 Collaboration],
  Phys.\ Rev.\  D {\bf 77} (2008) 052003;


\bibitem{E14}  http://koto.kek.jp/



\bibitem{Na62}  http://na62.web.cern.ch/na62/




\bibitem{Seidel:2004jh}
  D.~Seidel,
  Phys.\ Rev.\ D {\bf 70} (2004) 094038
  [hep-ph/0403185].


\end{thebibliography}
\end{document}